\documentstyle[12pt,titlepage]{article}
\input epsfig.sty

\font\grande=cmr9.5 scaled \magstep4
\font\medio=cmr9.5 scaled \magstep2
\outer\def\beginsection#1\par{\medbreak\bigskip
      \message{#1}\leftline{\bf#1}\nobreak\medskip
\vskip-\parskip
      \noindent}

\def\laq{\raise 0.4ex\hbox{$<$}\kern -0.8em\lower 0.62
ex\hbox{$\sim$}}
\def\gaq{\raise 0.4ex\hbox{$>$}\kern -0.7em\lower 0.62
ex\hbox{$\sim$}}

\begin{document}
\bibliographystyle {unsrt}

\titlepage

\begin{flushright}
CERN-PH-TH/2006-109
\end{flushright}

\vspace{15mm}
\begin{center}
{\grande Tight coupling expansion}\\
\vspace{3mm}
{\grande and fully inhomogeneous magnetic fields}\\
\vspace{15mm}
 Massimo Giovannini 
 \footnote{Electronic address: massimo.giovannini@cern.ch} \\
\vspace{6mm}

\vspace{0.3cm}
{{\sl Centro ``Enrico Fermi", Compendio del Viminale, Via 
Panisperna 89/A, 00184 Rome, Italy}}\\
\vspace{0.3cm}
{{\sl Department of Physics, Theory Division, CERN, 1211 Geneva 23, Switzerland}}
\vspace*{2cm}

\end{center}

\vskip 2cm
\centerline{\medio  Abstract}
The tight coupling expansion, appropriately generalized to include 
large-scale magnetic fields, allows the estimate of the brightness perturbations of CMB anisotropies for typical wavelengths that are larger than the Hubble radius after matter-radiation equality.  After discussing the basic features of the the pre-decoupling initial conditions in the presence of fully inhomogeneous magnetic fields, the tight coupling expansion is studied both analytically and numerically. From the requirement that the amplitudes and phases of Sakharov oscillations are (predominantly) adiabatic 
and from the inferred  value of the plateau in the temperature autocorrelation, the 
effects of the magnetized contribution can be systematically investigated and constrained.
\noindent

\vspace{5mm}

\vfill
\newpage
\renewcommand{\theequation}{1.\arabic{equation}}
\section{Introduction}
\setcounter{equation}{0}
A distinctive signature of pre-decoupling (adiabatic) initial conditions is that the temperature autocorrelation has the first (Doppler) peak for $\ell_{\rm d}\simeq 220$ \cite{wmap1,wmap2,wmap3}.
For the same set of initial conditions, the cross-correlation power spectrum between temperature and polarization will have the first (anticorrelation) peak for $\ell_{\rm c} \simeq 3 \,\ell_{\rm d}/4 < \ell_{\rm d} $ \cite{wmap1}. 

Defining as $k$ the comoving wave-number and as $c_{\rm sb}$ the sound speed of the baryon-photon system, the large-scale contribution to the temperature autocorrelation oscillates as a $\cos{(k c_{\rm sb } \tau_{\rm dec} )}$, while the cross-correlation oscillates as $\sin{( 2 k c_{\rm sb } \tau_{\rm dec}) }$.  The 
first (compressional) peak of the temperature autocorrelation will then 
correspond to $ k c_{\rm sb } \tau_{\rm dec}\sim \pi$ (i.e. $\ell_{\rm d} \simeq 220$),
while  the first peak of the cross-correlation 
will arise for $k c_{\rm sb } \tau_{\rm dec}\sim 3\pi/4$ (i.e., approximately, $\ell_{\rm c} \simeq 3\, \ell_{\rm d}/4$).
This result can be obtained analytically by working to first-order in the tight-coupling expansion \cite{TC1,TC2,TC3}. 
In this framework, the key assumption is that the initial conditions of the Einstein-Boltzmann hierarchy are characterized, prior to matter-radiation equality, by a solution that is, at least predominantly, adiabatic. 
A corollary to the previous statements is that the peak in the cross-correlation 
occurs for typical wavelengths that are larger than the Hubble radius at the 
the time of photon decoupling.

In a previous paper \cite{mg1} it has been shown that large-scale magnetic fields, present prior to  equality affect the CMB initial conditions. 
If only the adiabatic mode is present, the fully inhomogeneous magnetic fields contribute to the Sachs-Wolfe plateau as a subdominant non-adiabatic component that may be either correlated or uncorrelated  with the adiabatic mode. If the pre-decoupling  initial conditions are given by a mixture 
of adiabatic and non-adiabatic modes, the magnetized contribution may also mix with the isocurvature modes so that the total number of parameters defining 
the initial conditions increases.
It is desirable to complement and extend the analysis of \cite{mg1} by scrutinizing 
the impact of pre-decoupling magnetic fields on  intermediate scales.
Within the tight coupling expansion, appropriately generalized to include the effects 
of fully inhomogeneous magnetic fields, the fluctuations in the brightness induced by the
(adiabatic and non-adiabatic) scalar modes can be estimated. 

The effects of magnetic fields on the scalar brightness perturbations are typically 
neglected. To give an example, let us consider how the impact of large-scale 
magnetic fields is usually computed, for instance, in the Faraday effect. 
This case is interesting, since, the rotation of the CMB polarization plane 
indeed presupposes that the magnetic fields do not affect the process of 
formation of polarization. 
The influence of large-scale magnetic fields on CMB polarization 
can  be separated, for illustration, into two physically distinct effects.
Prior to decoupling, gravitating 
magnetic fields modify the initial conditions of CMB anisotropies and 
affect the evolution equations of the photon-baryon system. Therefore,
the induced degree of polarization will be sensitive to  
the presence of magnetic fields whose role will be, at this stage, to 
modify the features of the adiabatic initial conditions. This is the first effect.
Once the polarization is formed, the polarization 
plane of the CMB may be Faraday rotated \cite{F1,F2}. From this second effect interesting bounds on the magnetic field intensity 
are usually derived.

This philosophy motivated various independent studies 
on the possible Faraday rotation of the CMB polarization 
degree \cite{F3,F4,F5,F6}. In some cases the magnetic field 
has been assumed just uniform \cite{F3}, but there are also 
calculations analyzing the situation where the magnetic field 
is fully inhomogeneous \cite{F4,F5}.  In spite of the
relevant technical differences, the common assumption of the mentioned
investigations is that the magnetic field is negligible in 
the process of formation of CMB polarization. The latter assumption can be rephrased by saying that the 
initial conditions of the Einstein-Boltzmann hierarchy are 
set in such a way that the magnetic fields are absent. 
In the absence of large-scale magnetic fields the only  source of polarization 
arises to first-order in the tight coupling expansion. 
The magnitude of the 
obtained contribution is just a consequence of the adiabaticity of the solution.  
If large-scale magnetic fields are consistently 
introduced before matter-radiation equality, the 
usual adiabatic (as well as non-adiabatic) initial conditions of the 
Einstein-Boltzmann hierarchy are modified and, consequently, 
also the degree of polarization of the CMB is affected.

This example motivates a systematic analysis of the impact of magnetized 
initial conditions on the brightness perturbations of the radiation field. 
As anticipated above, this calculation will be conducted within the tight-coupling 
approximation which is known to be rather effective in obtaining the brightness 
perturbations for large and intermediate scales.  This analysis has never 
been done before, to the best of our knowledge. A related motivation is that, in the 
absence of magnetic fields, the tight coupling approximation is employed  in 
numerical Boltzmann codes at early times to avoid the integration of stiff
Euler equations. The results of the present 
investigation allow to use the same strategy also in the presence 
of fully inhomogenous magnetic fields.

Several aspects of CMB physics are germane to large-scale 
magnetization (see \cite{r1,r2} for two topical review articles). 
The astrophysical evidence for large-scale magnetization can 
 be summarized as follows. Large-scale magnetic fields are an important element of the 
physics of the interstellar medium and have been measured, through 
various techniques in galaxies  \cite{b1,b2} and  in rich clusters \cite{l1,l2,l3,l4}.
The determination of large-scale magnetic fields associated 
with loosely gravitationally bound systems (like superclusters) 
is still debatable \cite{k1,k2,k3}, but, nonetheless, extremely intriguing.
As far as galactic magnetic fields are concerned, there is the hope that 
in the future all sky survey of Faraday rotation (achievable through 
the Square Kilometer Array [SKA] \cite{b2,l1})  will allow a Faraday "tomography"
of the Milky Way addressing the dark corners (number of reversals, 
pitch angle) of the morphological 
structure of the closest large-scale magnetic field we observe through 
different surveys of pulsars such as the Parkes and the 
southern galactic plane surveys (see  \cite{han,parkes,sgps} and references therein). 
There is the hope, in the present framework, that the present-day coherent field in spiral galaxies 
may keep the "memory" of the structure of the initial seed field (see \cite{b1}
for a review of this long-standing problem).
The morphological features of the magnetic fields profiles in clusters 
are  less understood than in the case of galaxies. However, 
attempts have been made in constraining the radial profiles of magnetic 
fields in clusters \cite{l2,l3}.  
Determinations of intracluster magnetic fields through statistical 
samples of rich clusters \cite{l4} seem to suggest that (ordered?) intra-cluster 
magnetic fields may be as large as the $\mu {\rm G}$ (see, for instance, 
\cite{magnetized} for a review on the features of our magnetized Universe).

Compressional amplification (taking place during the gravitational collapse 
of the protogalaxy) allows to connect the observed field to a protogalactic field, 
present prior to gravitational collapse, between  $0.1$ nG and $0.01$ nG.
It is plausible, within the dynamo hypothesis, that the protogalactic 
fields could even be much smaller than the nG. In recent years, a lot 
of progress has been made both in the context of small- and large-scale 
dynamos i.e. the process by which the kinetic energy of the protogalactic plasma is transferred, 
by differential rotation, into magnetic energy. 
This progress \cite{rev1} (see also \cite{rev2} for an introduction 
to astrophysical dynamos) has been driven both by the higher resolution 
of numerical simulations and by the improvement in the 
understanding of the largest magnetized system that is close to us, i.e. 
the sun. In contrast  with 
the previous lore, it is now clear that the dynamo action demands a change 
in the topology of the flux lines. As a consequence, large-scale 
dynamos should also produce small-scale helical fields that quench 
(i.e. prematurely saturate) the $\alpha$ effect \cite{rev1} (see also \cite{rev3}).
The morphology and strength of magnetic fields in clusters may be related 
to the way the dynamo effect saturates.

The simplest way to understand why large-scale magnetic fields 
may be relevant for CMB physics is to think of a protogalactic 
magnetic field in the nG range. If this field is naively blue-shifted 
at the decoupling epoch, i.e. at a redshift $z_{\rm dec} \simeq 1100$
its strength could be as large as the mG, i.e. roughly six orders of magnitude 
larger. In the past ten years, indeed a lot of work has been 
done with the purpose of constraining large-scale magnetic fields 
using CMB physics. Historically the first type of configurations 
used in this type of exercise have been weakly uniform magnetic fields 
that would partially break the isotropy of the background geometry \cite{r2}. 
It was than realized that this assumption induces a correlation of the $a_{\ell -1,m}$ and 
$a_{\ell +1,m}$ multipole coefficients arising in the angular power spectrum 
(see \cite{unif1,unif2,unif3} and references therein). From this 
observation, uniform magnetic fields can be constrained.

Another (complementary) possibility is that magnetic fields are fully 
inhomogeneous. In this case the magnetic fields do not break the spatial 
isotropy of the background but they can affect virtually all observables relevant 
in the theory of CMB anisotropies. 
In particular, in a series of papers, 
Barrow and Subramanian \cite{bs1,bs2,bs3} pointed out 
possible effects of magnetic fields on the induced vector CMB anisotropies 
as well as on the polarization power spectra induced by the same modes.
Within a slightly different approach the vector modes (as well as the tensor 
modes) have been discussed in \cite{m1}.
More recently \cite{lw1} a full numerical analysis of the vector and of the 
tensor modes induced by fully inhomogeneous magnetic fields has been 
presented.  In \cite{rat1} the signatures of magnetic helicity have been 
studied always in the perspective of the vector modes. 
Finally, in a series of papers, the evolution of weakly inhomogeneous 
magnetic fields has been discussed within the covariant formalism 
and in the presence of a uniform magnetic background \cite{ts2}.

In spite of the fact that vector modes are probably one of the characteristic 
signatures of large-scale magnetic fields, scalar modes are also induced 
\cite{mg1,mg2}. 
This statement can be understood by noticing that the magnetic energy 
density, the magnetic anisotropic stress and the magnetic pressure 
contribute, respectively, to the Hamiltonian and momentum constraint and to the dynamical evolution of the curvature perturbations. Furthermore, within the magnetohydrodynamical approximation, the divergence 
of the Lorentz force contributes to the evolution equation of the 
divergence-full part of the baryon peculiar velocity.  

What is missing, at the moment, is the analysis of the brightness 
perturbations induced by magnetized initial conditions. As anticipated, one 
of the purposes of the present paper will exactly be to bridge fill this 
gap and to allow an approximate evaluation of the brightness 
perturbations, also at intermediate scales,  
as a function of the spectral properties of fully inhomogeneous 
magnetic fields.

The present paper is then organized as follows. 
In Section 2 the initial condtions of the Einstein-Boltzmann
hierarchy will be discussed for typical wavelengths 
larger than the Hubble radius prior to equality. In Section 3 
the tight-coupling equations will be derived and discussed 
analytically. The numerical treatment of the system is presented 
in Section 4. Section 5 contains our concluding remarks.

\renewcommand{\theequation}{2.\arabic{equation}}
\section{Adiabatic and non-adiabatic modes}
\setcounter{equation}{0}

\subsection{Zeroth-order tight-coupling approximation}

By combining the Thompson cross section (defined, in the following, as $\sigma_{\rm T}$) with the ionization fraction $x_{\rm e}$ and with the electron density $n_{\rm e}$,
the differential optical depth is given by
 $ \kappa' = n_{\rm e} x_{\rm e} \sigma_{\rm T} a/a_{0}$.   Since 
 prior to decoupling the photon mean free path is much smaller 
 than the Hubble radius at the corresponding epoch, 
  the inverse of $\kappa'$ can be used as an expansion parameter.  This is the 
  basic idea of the tight coupling approximation.
 The lowest order, the
tight-coupling approximation implies $1/{\kappa}' \to 0$ (or, equivalently,
that $\sigma_{\rm T}\to \infty$). 
In this limit the divergences of the peculiar velocities of photons (i.e. $\theta_{\gamma}$) and baryons (i.e. $\theta_{\rm b}$) are driven to a common value i.e. $\theta_{\gamma} \simeq \theta_{\rm b} = \theta_{\gamma{\rm b}}$.  In the following, the consequences 
of the zeroth-order in the tight coupling approximation will be investigated 
in the presence of fully inhomogeneous magnetic fields.
Consider, therefore, the case where 
large-scale magnetic fields are fully inhomogeneous 
and characterized, in Fourier space,by the following two-point function
\begin{equation}
\langle B_{i}(\vec{k},\tau) B^{j}(\vec{p},\tau) \rangle =  \frac{2\pi^2}{k^3} \,P_{i}^{j}(k)\, P_{\rm B}(k)\, \delta^{(3)}(\vec{k} + \vec{p}),
\label{Bcorr}
\end{equation}
where 
\begin{equation}
P_{i}^{j}(k) = \biggl(\delta_{i}^{j} - \frac{k_{i}k^{j}}{k^2} \biggr),\qquad P_{\rm B}(k) = A_{\rm B} 
\biggl(\frac{k}{k_{\rm p}}\biggr)^{\varepsilon};
\label{MPS}
\end{equation}
In Eq. (\ref{MPS}) $k_{\rm p}$ is the "pivot" scale at which the 
spectra are normalized (see also, below, Eq. (\ref{RSP})).
Since the magnetic fields are fully inhomogeneous the line element 
of the background geometry will not be anisotropic \footnote{If the expansion is 
anisotropic because of a uniform magnetic field oriented along a specific direction, as in the 
celebrated Zeldovich models \cite{zm,zm1}, the angular power spectrum does not depend solely 
upon $\ell$ but also upon $m$. These situations can be successfully constrained 
through WMAP data (see \cite{unif1,unif2,unif3}) and references therein.} (as it happens, for 
instance, in magnetized Bianchi-type backgrounds) and it 
can be parametrized as
\begin{equation}
ds^2 = a^2(\tau) [ d\tau^2 - d \vec{x}^2].
\label{met1}
\end{equation}
Consequently, 
the evolution equations of the background geometry will be solely 
determined by the total pressure and energy density of the system as 
\begin{equation}
{\mathcal   H}^2 = \frac{8\pi G}{3} a^2 \rho_{\rm t},\qquad 
{\mathcal   H}^2 - {\mathcal   H}' = 4\pi G a^2 (\rho_{\rm t} + p_{\rm t}),\qquad 
\rho_{\rm t}' + 3 {\mathcal   H} (\rho_{\rm t} + p_{\rm t})=0,
\label{FL}
\end{equation}
where ${\mathcal H} = a'/a$ and the prime denotes a derivation 
with respect to the conformal time coordinate $\tau$.
Denoting the longitudinal fluctuations of the metric (\ref{met1}) 
as\footnote{In Eq. (\ref{longfl})
and in what follows, $\delta_{\rm s}(...)$ denotes the scalar fluctuation of the corresponding quantity.}
\begin{equation}
\delta_{\rm s} g_{00} = 2 a^2 \phi, 
\qquad \delta_{\rm s} g_{ij} = 2 a^2 \psi \delta_{ij},
\label{longfl}
\end{equation} 
and selecting the conformally Newtonian gauge \cite{long1,long2}, 
the evolution equations of photons and baryons can be obtained in terms 
of the density contrasts (i.e. $\delta_{\gamma}$ and $\delta_{\rm b}$) and in terms 
of the divergence of the peculiar velocities (i.e. $\theta_{\gamma}$ and $\theta_{\rm b}$):
\begin{eqnarray}
&& \delta_{\gamma}' =  4  \psi' - \frac{4}{3} \theta_{\gamma},\qquad 
\theta_{\gamma}'      =  -\frac{1}{4} \nabla^2 \delta_{\gamma} - \nabla^2 \phi + \kappa' (\theta_{\rm b} - \theta_{\gamma}),
\label{gamma1}\\
&& \delta_{\rm b}'    =  3 \psi' - \theta_{\rm b}, \qquad \theta_{\rm b}' = - {\mathcal   H} \theta_{\rm b} -\nabla^2 \phi 
+ \frac{\vec{\nabla}\cdot[\vec{J} \times \vec{B}]}{a^4 \rho_{\rm b}}
+\frac{4}{3} \frac{\rho_{\gamma}}{\rho_{\rm b}} \kappa' (\theta_{\gamma} -\theta_{\rm b}).
\label{b1}
\end{eqnarray}
The term $\vec{J}\times \vec{B}$ parametrizes the Lorentz force 
contribution as it arises in a conducting plasma which is, however, globally neutral (see, for instance, \cite{MHD1,MHD2}). In this situation, typical of magnetohydrodynamics (MHD), the possible contribution 
of electric fields is suppressed by inverse powers of the conductivity as it will be discussed later.
By subtracting the two equations for the velocities introduced in
 (\ref{gamma1}) and (\ref{b1}), the combination
 $(\theta_{\gamma} - \theta_{\rm b})$ obeys 
 \begin{equation}
 (\theta_{\gamma} - \theta_{\rm b})' + \kappa' \biggl( 1 + \frac{4}{3} \frac{\rho_{\gamma}}{\rho_{\rm b}}\biggr) (\theta_{\gamma} - \theta_{\rm b})= 
- \frac{1}{4} \nabla^2\delta_{\gamma} + {\mathcal   H} \theta_{\rm b} +
 \frac{ \vec{\nabla} \cdot[\vec{J} \times \vec{B}]}{a^4 \rho_{\rm b}}.
\label{TC1}
\end{equation}
The term at the right hand side of Eq. (\ref{TC1}) can be 
viewed as a source term for the  evolution of the difference 
between the divergences of the photon and baryon peculiar 
velocities. In spite of the magnitude of the source term at the right hand side 
of Eq. (\ref{TC1}), $(\theta_{\gamma} - \theta_{\rm b})$ will quickly 
be driven to zero in the limit $\sigma_{\rm T} \to \infty$.
 Consequently, 
even if pre-decoupling magnetic fields are present, 
the lowest order tight coupling expansion still implies that 
$\theta_{\gamma}\to \theta_{\rm b}$ as expected.

In the tightly coupled limit, the evolution equation  of 
$\theta_{\gamma{\rm b}}= \theta_{\gamma} = \theta_{\rm b}$ can then
be swiftly obtained by summing up the equations for the velocity 
fields in such a way that the scattering terms cancel. The result 
of this procedure is, in Fourier space,
\begin{equation}
\theta_{\gamma{\rm b}}' + \frac{{\mathcal   H} R_{\rm b}}{R_{\rm b} + 1} \theta_{\gamma{\rm b}} = \frac{k^2}{4 ( 1 + R_{\rm b})}\delta_{\gamma} + k^2 \phi + \frac{k^2}{4 ( 1 + R_{\rm b})} ( \Omega_{\rm B} - 4 \sigma_{\rm B}),
\label{TC2}
\end{equation}
where $R_{\rm b} = (3/4) \rho_{\rm b}/\rho_{\gamma}$.  
For numerical purposes (see Section 4) the late-time cosmological parameters will be fixed, for a spatially flat Universe, as
\footnote{The values of the cosmological parameters introduced in Eq. (\ref{par})
are compatible with the ones estimated from WMAP-3 \cite{wmap2,wmap3} 
in combination with the ``Gold" sample of SNIa \cite{riess} consisting of $157$ 
supernovae (the furthest being at redshift z = 1.75). We are aware of the fact 
that WMAP-3 data alone seem to favour a slightly smaller value of $\omega_{\rm m}$ 
(i.e. $0.126$). Moreover, WMAP-3 data may also have slightly 
different implications if combined with supernovae of the SNLS project \cite{astier}.
The values given in Eq. (\ref{par}) will just be used for a realistic numerical illustration
of the methods developed in the present investigation.}
\begin{equation}
\omega_{\gamma} = 2.47\times 10^{-5},\qquad \omega_{\rm b} = 0.023,\qquad 
\omega_{\rm c} = 0.111, \qquad \omega_{\rm m} = \omega_{\rm b} + \omega_{\rm c},
\label{par}
\end{equation} 
where $\omega_{X} = h^2 \Omega_{X}$ and 
$\Omega_{\Lambda}= 1- \Omega_{\rm m}$; the present 
value of the Hubble parameter $H_{0}$ will be fixed, for numerical 
estimates,  to $73$ in units 
of ${\rm km}/({\rm sec}\, {\rm Mpc})$.  In terms of the fiducial 
set of parameters of Eq. (\ref{par}) the baryon to photon ratio 
appearing in Eq. (\ref{TC2}) becomes 
\begin{equation}
R_{\rm b} = \biggl(\frac{698}{z +1}\biggr) \biggl(\frac{\omega_{\rm b}}{0.023}\biggr) \biggl(\frac{\omega_{\gamma}}{2.47\times 10^{-5}}\biggr)^{-1},
\label{RB}
\end{equation}
where $z + 1 = a_{0}/a$ is the redshift (for numerical purposes 
the scale factor will be normalized in such a way that $a_{0} = 1$; see 
below, Eq. (\ref{SF})).
In Eq. (\ref{TC2}) we traded the MHD current $\vec{J}$ (appearing in the 
second equation of (\ref{b1})) for a combination 
of the magnetic energy density and of the anisotropic stress. More specifically, 
it can be easily shown that 
\begin{equation}
\frac{3}{4} \frac{\vec{\nabla} \cdot [\vec{J} \times \vec{B}]}{a^4 \rho_{\gamma}}= \nabla^2 \sigma_{\rm B} - \frac{1}{4} \nabla^2 \Omega_{\rm B},\qquad
\Omega_{\rm B} = \frac{ \delta \rho_{\rm B}}{ \rho_{\gamma}}, \qquad 
\delta\rho_{\rm B} = \frac{B^2}{8\pi a^4}.
\label{an1}
\end{equation}
Equation (\ref{an1}) holds since, in MHD, the total Ohmic current $\vec{J}$ 
is solenoidal (i.e. $\vec{\nabla}\cdot \vec{J} =0$) and given 
by $ 4 \pi \vec{J} = \vec{\nabla}\times \vec{B}$ \cite{MHD1,MHD2} (see also the appendix of \cite{magnetized}). In addition, the function 
$\sigma_{\rm B}$ is nothing but a parametrization of the anisotropic stress 
that appears in the spatial part of magnetic energy-momentum tensor, i.e.
\begin{equation}
\delta_{\rm s} {\mathcal   T}_{i}^{j} = - \delta p_{\rm B} + \tilde{\Pi}_{i}^{j},\qquad \partial_{i}\partial^{j}\tilde{\Pi}_{i}^{j} = (p_{\gamma} +\rho_{\gamma}) \nabla^2 \sigma_{\rm B}, 
\label{dt0}
\end{equation}
where 
\begin{equation}
\delta p_{\rm B} = \frac{\delta \rho_{\rm B}}{3}, \qquad 
\tilde{\Pi}_{i}^{j} = 
\frac{1}{4 \pi a^4}\biggl(B_{i} B^{j} - \frac{B^2}{3} \delta_{i}^{j}\biggr).
\label{dt0a}
\end{equation}
In various context the force-free approximation is often employed 
\cite{olesen1, olesen2} (see also 
\cite{magnetized} and references therein). Such an approximation amounts, in practice, to set 
$(\vec{\nabla}\times \vec{B}) \times \vec{B} \to 0$. In this limit, from Eq. (\ref{an1}) 
$\sigma_{\rm B} \to \Omega_{\rm B}/4$. In the present framework this 
assumption shall not be invoked.  

According to Eq. (\ref{an1}), the leading contribution of fully inhomogeneous 
magnetic fields to the scalar modes of the geometry is 
given by the magnetic energy density and by the divergence of the 
MHD Lorentz force. These two quantities, affect the perturbed Einstein equations
 and, in particular, the Hamiltonian constraint 
\begin{equation}
\nabla^2 \psi - 3 {\mathcal   H} ({\mathcal   H} \phi + \psi') = 4\pi G a^2 [ \delta \rho_{\rm t} 
+ \delta_{\rm s} \rho_{\rm B}], 
\label{ham0}
\end{equation}
and the total (scalar) anisotropic stress 
\begin{equation}
\nabla^4 (\phi - \psi) =  12 \pi G a^2 [ (p_{\nu} + \rho_{\nu} ) \nabla^2 \sigma_{\nu} +  
( p_{\gamma} + \rho_{\gamma}) \nabla^2 \sigma_{\rm B}].
\label{anis2}
\end{equation}
Equations (\ref{ham0}) and (\ref{anis2}) follow, respectively, 
 from the $(00)$ and  $(i\neq j)$ components of the perturbed 
 Einstein equations written in the conformally Newtonian 
 coordinate system; the quantity  $\delta_{\rm s} \rho_{\rm t}$ 
 is the (total) scalar fluctuation of the energy-density 
 of the fluid sources.
 It is relevant to remark that, in Eq. 
 (\ref{anis2}), on top of the anisotropic stress of the magnetic field, 
 the only source of the anisotropic stress of the fluid 
 is provided by the neutrinos that are collisionless after weak interactions 
 fall out of thermal equilibrium for typical temperatures of the order 
 of $0.8$ MeV. In analogy with Eq. (\ref{dt0}), the anisotropic stress 
 of the neutrinos has been parametrized as  
 \begin{equation}
 \delta_{\rm s} T_{i}^{j} = - \delta_{\rm s} p_{\rm t} + \Pi_{i}^{j},\qquad \partial_{i}\partial^{j}\Pi_{i}^{j} = (p_{\nu} +\rho_{\nu}) \nabla^2 \sigma_{\nu}.
 \label{dt1}
 \end{equation}
 The quantity $\delta_{\rm s} p_{\rm t}$ is the fluctuation of the total pressure.
The last constraint on the dynamical evolution 
us derived from the $(0i)$ components of the Einstein equations and it is 
\begin{equation}
\nabla^2 ( {\mathcal   H} \phi + \psi') = - 4\pi G a^2 \biggl[ 
( p_{\rm t} + \rho_{\rm t}) \theta_{\rm t}  
+ \frac{ \vec{\nabla} \cdot ( \vec{E}\times \vec{B})}{4 \pi a^4}\biggr].
\label{mom0}
\end{equation}
The second term at the right hand side of Eq. (\ref{mom0}) is 
the divergence of the Poynting vector. In MHD the Ohmic electric field 
is subleading and, in particular, from the MHD expression 
of the Ohm law 
\begin{equation}
\vec{E} \times \vec{B} = \frac{(\vec{\nabla}\times \vec{B})\times \vec{B}}{4\pi\sigma}.
\end{equation}
Since the Universe, prior to decoupling, is a very good conductor, the 
the ideal MHD limit \cite{MHD2} can be safely adopted in the first approximation (see also \cite{bir}); thus 
for $\sigma \to \infty$ the contribution of the Poynting vector 
vanishes. In any case, even if $\sigma $ would be finite but large, the second 
term is suppressed in comparison with the first term at the right hand side of Eq. (\ref{mom0}); such a term is given by the sum of the divergence of the peculiar 
velocity of each fluid component weighted by the specific enthalphy (i.e. 
$(p_{\rm a} + \rho_{\rm a})$ for a generic a-fluid):
\begin{equation}
(p_{\rm t} + \rho_{\rm t}) \theta_{\rm t}  = \sum_{a} (p_{\rm a} + \rho_{\rm a}) \theta_{\rm a},
\end{equation}
or, in more explicit terms, 
\begin{equation}
(p_{\rm t} + \rho_{\rm t}) \theta_{\rm t} \equiv \frac{4}{3} \rho_{\nu} \theta_{\nu}
+ \frac{4}{3} \rho_{\gamma} \theta_{\gamma} + \rho_{\rm c} \theta_{\rm c}
+ \rho_{\rm b} \theta_{\rm b} = \frac{4}{3} \rho_{\nu} \theta_{\nu} + \rho_{\rm c} \
+ \frac{4}{3} \rho_{\gamma}( 1 + R_{\rm b}) \theta_{\gamma{\rm b}}.
\label{TOTV}
\end{equation}
The second equality in Eq. (\ref{TOTV})  follows from the zeroth order in the tight-coupling expansion and from the definition of the baryon to photon ratio
(\ref{RB}). Every regular solution to lowest order in the tight coupling expansion has to satisfy the momentum constraint of Eq. (\ref{mom0}). Such a  
requirement implies, in particular, a restriction on the form of the 
peculiar velocity fields for comoving wave-numbers larger 
than the Hubble radius (i.e. $k\tau \ll 1$).

The evolution of the cold dark matter (CDM)
 component only feels the presence of the 
magnetic field through the modifications induced by $\Omega_{\rm B}$ 
and $\sigma_{\rm B}$ on  of the two conformally 
Newtonian potentials, i.e. $\phi$ and $\psi$:
\begin{equation}
\delta_{\rm c}' = 3 \psi' - \theta_{\rm c},\qquad \theta_{\rm c}' + {\mathcal   H} \theta_{\rm c} = k^2 \phi.
\label{CDM0}
\end{equation}
A similar situation occurs for the evolution of the neutrino component 
where, however,  the effect of the magnetic field enters the not only the 
Newtonian potentials but also the anisotropic 
stress that is constrained from Eq. (\ref{anis2}). The relevant equations for the neutrinos
 are then, in Fourier space,
\begin{equation}
\delta_{\nu}' = 4\psi' - \frac{4}{3} \theta_{\nu}, \qquad \theta_{\nu}' = \frac{k^2}{4} \delta_{\nu} - k^2 \sigma_{\nu} + k^2 \phi, \qquad 
\sigma_{\nu}' = \frac{4}{15} \theta_{\nu}  - \frac{3}{10} k {\mathcal   F}_{\nu\,3}.
\label{NU0}
\end{equation}
The quantity ${\mathcal   F}_{\nu\,3}$ appearing in Eq. (\ref{NU0}) is the octupole 
of the (perturbed) neutrino phase space distribution while $\sigma_{\nu}$ 
(i.e. 
the anisotropic stress of the neutrinos introduced in Eqs. (\ref{anis2}) and (\ref{dt1})) is related to the quadrupole moment
of ${\mathcal   F}_{\nu}$ as $\sigma_{\nu} = {\mathcal   F}_{\nu\,2}/2$. The 
rationale for the inclusion of ${\mathcal   F}_{\nu\,3}$ in the fluid description 
is that  neutrinos are collisionless for temperature smaller 
than $0.8$ MeV when the Universe is still deep into the radiation-dominated regime. 
In this regime the neutrinos
 do not interact, and, consequently, 
 they should be rather described by the collisionless  Boltzmann 
hierarchy of the (perturbed) neutrino phase space distribution. 
To address this problem,  
the usual approach is to use an improved description where 
on top of the monopole and dipole of the (perturbed) neutrino 
phase space distribution (connected, respectively, with $\delta_{\nu}$ and $\theta_{\nu}$), one also takes into account the quadrupole, i.e. $\sigma_{\nu}$ 
and sometimes (as in the case of specific non-adiabatic modes) 
also the octupole.

With all these ingredients in mind, the evolution of the longitudinal fluctuations of the geometry is determined from the trace spatial components of the perturbed Einstein equations, and can be written as 
\begin{equation}
\psi'' + {\mathcal   H}( \phi' + 2 \psi') + ({\mathcal   H}^2 + 2 {\mathcal   H}') \phi + \frac{1}{3} \nabla^2 ( \phi - \psi) = 4\pi G a^2 (\delta_{\rm s} p_{\rm t} 
+ \delta p_{\rm B}),
\label{psi}
\end{equation}
where $\delta_{\rm s} p_{\rm t}$ is the fluctuation of the total pressure introduced in Eq. (\ref{dt1}).
In the present situation it will be mandatory to set initial conditions for the
fluctuations of the geometry and for the various species of the plasma before radiation-matter equality but, also, after neutrino decoupling. 
In the conformally flat case described by Eq. (\ref{met1}), the scale factor that
interpolates between the radiation epoch and the matter-dominated epoch can 
be obtained by integrating  Eqs. (\ref{FL}): 
\begin{equation}
a(\tau) = a_{\rm eq} \biggl[ \biggl(\frac{\tau}{\tau_1}\biggr)^2 + 2 \biggl(\frac{\tau}{\tau_{1}}\biggr) \biggr], \qquad 1 + z_{\rm eq} = \frac{1}{a_{\rm eq}} = \frac{h^2 \Omega_{{\rm m}0}}{h^2 \Omega_{{\rm r}0}},
\label{SF}
\end{equation}
 where $\Omega_{{\rm m}0}$ and $\Omega_{{\rm r}0}$ are evaluated at the present time and the 
 scale factor is normalized in such a way that $a_0= 1$. In Eq. (\ref{SF}) $\tau_{1} = (2/H_{0}) 
 \sqrt{a_{\rm eq}/\Omega_{{\rm m}0}}$. In terms of $\tau_{1}$ the equality time is 
 \begin{equation}
 \tau_{\rm eq} = (\sqrt{2} -1) \tau_{1} = 119.07 \,\, \biggl( \frac{h^2 \Omega_{{\rm m}0}}{0.134}\biggr)^{-1}\,\, {\rm Mpc},
 \label{SF1}
 \end{equation}
 i.e.  $2 \tau_{\rm eq} \simeq \tau_{1}$.
 In this framework the total optical depth from the present to the critical recombination epoch, i.e. $ 800< z < 1200$
 can be approximated analytically,  as discussed in \cite{wh}.  By defining the redshift of decoupling as the one where 
 the total optical depth is of order 1, i.e. $\kappa(z_{\rm dec},0) \simeq 1$, we will have, approximately 
 \begin{equation}
 z_{\rm dec}  \simeq 1139 \biggl(\frac{ \Omega_{\rm b}}{0.0431}\biggr)^{- \alpha_{1}}, \,\,\,\,\,\,\,\,\alpha_{1} = \frac{0.0268}{0.6462 + 0.1125\ln{(\Omega_{\rm b}/0.0431)}},
 \label{zdec}
\end{equation}
 where consistently with Eq. (\ref{par}), $h = 0.73$.  From Eqs. (\ref{zdec}) and (\ref{SF}) it follows 
 that for $ 1100 \leq z_{\rm dec} \leq 1139$, $ 275\,\, {\mathrm Mpc} \leq \tau_{\rm dec} 
 \leq 285\,\, {\mathrm Mpc}$.
 
\subsection{Magnetized adiabatic mode}
In the limit $\tau/\tau_{1} \ll 1$, the solution of the whole system of governing equations corresponding to the adiabatic mode can be obtained 
as an expansion in powers of $k\tau$. When the given 
wavelength is larger than the Hubble radius before equality (i.e. 
$k\tau< 1$ for  $\tau < \tau_{\rm eq}$ )the solution of the coupled system 
introduced in the previous subsection can be written  as\footnote{In the following, 
all the perturbed quantities 
will be defined in Fourier space and will depend on the comoving wave-number $k$.} 
\begin{eqnarray}
&& \delta_{\nu} = \delta_{\gamma} = - 2 \phi_{*} - R_{\gamma} \Omega_{\rm B},
\qquad \delta_{\rm c} = \delta_{\rm b} = - \frac{3}{2} \phi_{*} - \frac{3}{4} R_{\gamma} \Omega_{\rm B}
\label{DCad}\\
&& \theta_{\rm c} = \frac{k^2 \tau}{2} \phi_{*},\qquad \theta_{\nu}  = \frac{k^2 \tau}{2}\biggl[ \phi_{*} - 
\frac{R_{\gamma} \Omega_{\rm B}}{2} + 2 \frac{R_{\gamma}}{R_{\nu}} \sigma_{\rm B} \biggr], 
\label{THad}\\
&&\theta_{\gamma{\rm b}} = \frac{k^2 \tau}{2} \biggl[ \phi_{*} + 
\frac{R_{\nu} \Omega_{\rm B}}{2} - 2 \sigma_{\rm B}\biggr],
\label{THgad}\\
&&\psi_{*} = \phi_{*} \biggl( 1 + \frac{2}{5} R_{\nu} \biggr) + \frac{R_{\gamma}}{5} [ 4 \sigma_{\rm B}   - R_{\nu}\Omega_{\rm B}], \qquad \sigma_{\nu} = - \frac{R_{\gamma}}{R_{\nu}} \sigma_{\nu} + \frac{k^2 \tau^2}{6 R_{\nu}} ( \psi_{*} - \phi_{*}),
\label{longad}
\end{eqnarray}
where the fractional contribution of photons to the radiation plasma, i.e. $R_{\gamma}$ has been introduced and it is related to $R_{\nu}$, i.e. 
the fractional contribution of massless neutrinos, as \footnote{In the present paper 
$h^2 \Omega_{\nu} = 1.68\times 10^{-5}$, i.e.,  recalling Eq. (\ref{par}), 
 $h^2 \Omega_{{\rm r}0} = h^2 \Omega_{\gamma} + h^2 \Omega_{\nu} = 
 4.15 \times 10^{-5}$. According to Eq. (\ref{SF}) this implies that $1 + z_{\rm eq} \simeq  3228.9 (\omega_{\rm m}/0.134)$. Possible mass
 terms for the neutrinos in the ${\rm meV}$ range will be neglected 
 here but can be introduced with a modification of the neutrino 
 Boltzmann hierarchy used to derive the relations reported in Eq. 
 (\ref{NU0}).} 
\begin{equation}
R_{\gamma} = 1 - R_{\nu}, \qquad R_{\nu} = \frac{r}{1 + r},\qquad r= \frac{7}{8} N_{\nu} \biggl(\frac{4}{11}\biggr)^{4/3} \equiv  0.681 \biggl(\frac{N_{\nu}}{3}\biggr).
\label{RNU}
\end{equation}
In Eq. (\ref{longad}) $\phi_{*}(k)$ and $\psi_{*}(k)$ denote the values of the Fourier modes of the longitudinal fluctuations of the geometry that are (approximately) constant 
in time. 
Equations (\ref{DCad}), (\ref{THad}) and (\ref{longad}) define a solution that is  adiabatic since the 
density contrasts in matter (i.e. $\delta_{\rm c}$ and $\delta_{\rm c}$) are exactly $3/4$ of the 
density contrasts in radiation (i.e. $\delta_{\gamma}$ and $\delta_{\nu}$).  
In the limit 
$\Omega_{\rm B} \to 0$ (vanishing magnetic energy density) and $\sigma_{\rm B}\to 0$ (vanishing magnetic anisotropic stress),  Eqs. (\ref{DCad}), (\ref{THad}) and (\ref{longad}) reproduce the standard adiabatic mode deep in the radiation dominated epoch \cite{mg2,mb}. From the second equation 
reported in (\ref{longad}) it is clear that the neutrino anisotropic stress is partially  counterbalanced
by the magnetic anisotropic stress. This aspect is entirely due to the 
presence of the magnetic anisotropic stress and it has been verified 
in different situations involving also entropic fluctuations \cite{mgn}.

In view of the forthcoming numerical applications, it is useful 
 to characterize the adiabatic mode in fully gauge-invariant terms. 
 In fact the quantity introduced so far are meaningful in the longitudinal gauge.
 The spectrum of curvature perturbations used as initial condition for the 
 numerical evaluation of the brightness perturbations is often introduced 
 \cite{wmap1,wmap2,wmap3} in terms of 
the curvature perturbations on comoving orthogonal hypersurfaces \cite{bst,bkp}, customarily denoted with 
${\mathcal   R}$. In the longitudinal gauge we have that this quantity is simply given by
\begin{equation}
 {\mathcal   R} = -\psi - \frac{{\mathcal   H} ({\mathcal   H} \phi + \psi')}{{\mathcal   H}^2 - {\mathcal   H}'}.
\label{calardef}
\end{equation}
In spite of the fact that ${\mathcal   R}$ is here defined in terms of the variables 
of the longitudinal gauge, its value is invariant under infinitesimal coordinate 
transformations.
In terms of ${\mathcal   R}$ the longitudinal fluctuations of the geometry entering the solution of Eqs.  (\ref{DCad}), (\ref{THad}) and (\ref{longad}) become:
\begin{eqnarray}
&&\phi_{*} = - \frac{10 {\mathcal R_{*}}}{ 15 + 4 R_{\nu}} - 
\frac{2 R_{\gamma}( 4 \sigma_{\rm B} - R_{\nu} \Omega_{\rm B})}{(15 + 4 R_{\nu})},
\nonumber\\
&&\psi_{*} = - \frac{ 2 ( 5 + 2 R_{\nu})}{15 + 4 R_{\nu}} {\mathcal   R}_{*} + \frac{R_{\gamma}}{15 + 4 R_{\nu}} [ 4  \sigma_{\rm B} - R_{\nu} \Omega_{\rm B}],
\label{longad2}
\end{eqnarray}
where ${\mathcal   R}_{*}(k)$ is the constant value of curvature perturbations 
prior to matter-radiation equality. Within the notations followed in this paper, 
the spectrum of ${\mathcal   R}_{*}(k)$ is defined as 
\begin{equation}
 \langle {\mathcal   R}_{*}(\vec{k}) {\mathcal   R}_{*}(\vec{p}) \rangle 
 = \frac{2\pi^2 }{k^3} 
{\mathcal  P}_{\mathcal   R}(k) \delta^{(3)}(\vec{k} + \vec{p}),\qquad 
{\mathcal  P}_{\mathcal   R}(k)=A_{{\mathcal   R}} \biggl(\frac{k}{k_{\rm p}}\biggr)^{n_{r} -1}.
\label{RSP}
\end{equation}
In Eq. (\ref{RSP})  $k_{\rm p}$ is the "pivot" scale (already introduced in Eq. (\ref{MPS})) at which the spectra are normalized. Typical choices of $k_{\rm p}$ range 
from $0.05\,\, {\rm Mpc}^{-1}$ down to $0.002 \,\, {\rm Mpc}^{-1}$
\cite{wmap1,wmap2,wmap3}.

Equations (\ref{Bcorr}) and (\ref{MPS}) allow to determine 
the spectra of $\Omega_{\rm B}(k)$ and $\sigma_{\rm B}(k)$. In the 
present notations 
\footnote{In the present paper the calculations are consistently conducted within the parametrization introduced 
in Eqs. (\ref{Bcorr}) and (\ref{MPS}). Different authors use slightly different conventions 
which can be, however, easily translated in the present parametrization. 
For instance, the magnetic spectral index defined as $n$ 
in \cite{r1,bs1,m1,lw1} correspond, in the present notations, to $\varepsilon - 3$.
In \cite{r1,bs1,m1,lw1} the factor $2\pi^2/k^3$ appearing in Eq. (\ref{Bcorr}) is omitted.
These definitions are clearly conventional but 
we regard as preferable the notation adopted here since it agrees with the usual
way of assigning the power spectrum of curvature perturbations in CMB physics. 
The other remark is that the magnetic fields considered here are not helical (see, for instance, 
\cite{rat1} and references therein). This is not a limitation since it can be 
easily appreciated that the helical contribution does not affect the scalar modes which are 
the main subject of the present investigation. }
\begin{equation}
\langle \Omega_{\rm B}(\vec{k}) \Omega_{\rm B}(\vec{p})\rangle = \frac{2\pi^2 }{k^3} 
{\mathcal   P}_{\Omega}(k) \delta^{(3)}(\vec{k} + \vec{p}),\qquad \langle \sigma_{\rm B} (\vec{k}) \sigma_{\rm B}(\vec{p}) \rangle = 
\frac{2 \pi^2}{k^3} {\mathcal   P}_{\sigma}(k) \delta^{(3)}(\vec{k} +\vec{p}),
\label{autoc}
\end{equation}
where 
\begin{equation}
{\mathcal   P}_{\Omega}(k) = {\mathcal   F}(\varepsilon) \overline{\Omega}_{{\rm B}\, L}^2 \biggl(\frac{k}{k_{L}}\biggr)^{2 \varepsilon},
\qquad
{\mathcal   P}_{\sigma}(k) = {\mathcal   G}(\varepsilon) \overline{\Omega}_{{\rm B}\, L}^2 \biggl(\frac{k}{k_{L}}\biggr)^{2 \varepsilon}.
\label{POM}
\end{equation}
The power spectra of Eq. (\ref{POM}) follow from the definitions 
of $\Omega_{\rm B}$ and $\sigma_{\rm B}$ by using the two-point function of the 
magnetic field, i.e. Eq. (\ref{Bcorr}). In fact, according to Eqs. (\ref{an1}) and (\ref{dt0a}), 
$\Omega_{\rm B}$ and $\sigma_{\rm B}$ are quadratic in the magnetic field intensity, and, therefore, lead to two mode-coupling integrals that have been estimated in
in the nearly scale-invariant limit (i.e. $\varepsilon < 1$) where the functions 
${\mathcal F}(\varepsilon)$ and ${\mathcal G}(\varepsilon)$ are determined to be 
\begin{equation}
{\mathcal   F}(\varepsilon) = \frac{4(6 - \varepsilon) ( 2 \pi)^{ 2 \varepsilon}}{\varepsilon ( 3 - 2 \varepsilon)
 \Gamma^2(\varepsilon/2)}, \qquad 
 {\mathcal   G}(\varepsilon) = 
 \frac{4(188 - 4 \varepsilon^2 - 66\varepsilon) (2\pi)^{2 \varepsilon}}{3 \varepsilon ( 3 - \varepsilon) ( 2 \varepsilon +1) \Gamma^2(\varepsilon/2)}.
 \label{FG}
 \end{equation}
 In the opposite limit (i.e. $\varepsilon \gg 1$) the mode-coupling integrals 
 arising in the expressions of $\Omega_{\rm B}$ and $\sigma_{\rm B}$ 
 are dominated by (small scale) diffusive effects ( see, for instance, 
 \cite{bs1,m1}) and the power spectra are sensitive to the specific 
 scale of the diffusion (Alfv\'en) damping. 
 In Eq. (\ref{POM}) the quantities
\begin{equation}
\overline{\Omega}_{{\rm B}\,\, L} = \frac{\rho_{{\rm B}\,L}}{\overline{\rho}_{\gamma}}, 
\qquad \rho_{{\rm B}\,\, L}=\frac{ B_{L}^2}{8\pi}, 
\qquad \overline{\rho}_{\gamma} = a^4(\tau)\rho_{\gamma}(\tau)
\label{OMC2}
\end{equation}
have been introduced. Notice that $B_{L}$ measures the 
magnetic field intensity smoothed over a comoving length-scale $L$.
In fact, by using a Gaussian window function $e^{ - k^2 L^2/2}$ for each 
magnetic field intensity, the (real space) magnetic autocorrelation 
function at two coincident spatial points is 
\begin{equation}
B_{L}^2 = \langle B_{i}(\vec{x},\tau) B^{i}(\vec{x},\tau) \rangle =  
A_{\rm B} \biggl(\frac{k_{L}}{k_{\rm p}}\biggr)^{\varepsilon} 
\Gamma\biggl(\frac{\varepsilon}{2}\biggr) (2\pi)^{-\varepsilon}
\label{AB}
\end{equation}
where $k_{L} = 2\pi/L$.  Equation (\ref{AB}) implies that the amplitude of the magnetic power spectrum 
appearing in Eqs. (\ref{Bcorr}) and (\ref{MPS}) 
can be directly characterized in terms of $B_{L}^2$ as 
\begin{equation}
A_{\rm B} = (2\pi)^{\varepsilon} \frac{B_{L}^2}{\Gamma(\varepsilon/2)} \biggl(\frac{k_{\rm p}}{k_{L}}\biggr)^{\varepsilon}.
\label{AtoB}
\end{equation}
It should be mentioned, incidentally, that nearly-scale invariant 
magnetic energy spectra can be achieved both in some class of inflationary models \cite{bharat} and in the case
of pre-big bang models \cite{pbb}. It is not our purpose here to endorse any specific model. The spirit of this investigation is more modest since we shall be content of understanding if large-scale magnetic fields can be consistently included in the picture of pre-decoupling physics and in the framework of  the usual techniques commonly employed for the description of scalar CMB anisotropies.

On top of the curvature fluctuations on comoving orthogonal hypersurfaces, another 
class of gauge-invariant quantities is related to the density contrasts both
total and partial (i.e. pertaining to each single species of the plasma).
The total density contrast on uniform density hypersurfaces
can be defined as a functional of the longitudinal degrees of freedom
\begin{equation}
\zeta = - \psi - \frac{{\mathcal   H} (\delta \rho_{\rm t} + \delta \rho_{\rm B})}{\rho_{\rm t}'},
\label{zetadef}
\end{equation}
coincides, for wavelengths larger than the Hubble radius, with ${\mathcal   R}$. In fact, using the definitions (\ref{calardef}) and (\ref{zetadef}) into the Hamiltonian 
constraint of Eq. (\ref{ham0}) it can be verified that 
\begin{equation}
\zeta = {\mathcal   R} + \frac{\nabla^2 \psi}{12 \pi G ( \rho_{\rm t} + p_{\rm t})}.
\end{equation}
Having introduced $\zeta$, related to the total density contrast, it is also useful to introduce the following 4 variables
\begin{equation}
\zeta_{\nu} = - \psi + \frac{\delta_{\nu}}{4}, \qquad \zeta_{\gamma} = - \psi + \frac{\delta_{\gamma}}{4}, \qquad 
\zeta_{\rm c} = - \psi + \frac{\delta_{\rm c}}{3},\qquad \zeta_{\rm b} = - \psi + \frac{\delta_{\rm b}}{3},
\label{zetapar}
\end{equation}
which are interpreted as the density contrasts for each independent 
fluid on uniform curvature hypersurfaces.

The evolution  for the density contrasts described by  Eqs. (\ref{gamma1}), (\ref{b1}), (\ref{CDM0}) and (\ref{NU0}) become, in terms of the variables 
introduced in (\ref{zetapar}), 
\begin{equation}
\zeta_{\gamma} '= - \frac{\theta_{\gamma{\rm b}}}{3},\qquad \zeta_{\nu}' = - \frac{\theta_{\nu}}{3}, \qquad 
\zeta_{\rm c}' = - \frac{\theta_{\rm c}}{3},\qquad \zeta_{{\rm b}}' = - \frac{\theta_{\gamma{\rm b}}}{3}.
\label{DCnew}
\end{equation}
Equations (\ref{zetapar}) and (\ref{DCnew}) are a useful tool for 
the discussion of the non-adiabatic solutions of the system as it will 
now be explicitly shown.

\subsection{Magnetized non-adiabatic modes}
The inclusion of fully inhomogeneous magnetic fields not only affects 
the adiabatic mode but also the other (non-adiabatic) modes whose physical features have been investigated in the conventional case
(see, for instance, \cite{mgn,cal,tur,h1,h2,h3} and references therein).
The adiabatic solution obtained in Eqs.  (\ref{DCad}), 
(\ref{THad}) and (\ref{longad}) implies,
in terms of the quantities defined in Eq. (\ref{zetapar}), that
\begin{equation}
\zeta_{\nu} = \zeta_{\gamma} = \zeta_{\rm c} = 
\zeta_{\rm b} = {\mathcal   R}_{*} - \frac{R_{\gamma}}{4} \Omega_{\rm B}.
\label{longad3}
\end{equation}
Equation (\ref{longad3}) is not accidental. 
It implies, in fact, that the entropy fluctuations 
of the mixture of fluids are vanishing. The entropy fluctuations are indeed 
defined as
\begin{equation}
{\mathcal   S}_{\rm i\,j} = - 3 (\zeta_{\rm i} - \zeta_{\rm j}),
\label{entrodef}
\end{equation}
where the indices run over the four components of the fluid defined previously.
Equation (\ref{longad3}) implies, therefore, 
 that the the entropy fluctuations vanish for the adiabatic mode.

The  pre-decoupling system also admit solutions where 
entropy fluctuations do not vanish. These modes are called non-adiabatic. 
If there are entropy fluctuations, the perturbation of the total pressure, appearing in Eqs. (\ref{dt1}) and (\ref{psi}), 
can be written as 
\begin{equation}
\delta_{\rm s} p_{\rm t} = c_{\rm s}^2 \delta \rho_{\rm t} + \delta p_{\rm nad}
\label{nad}
\end{equation}
where 
\begin{equation}
\delta p_{\rm nad} = \frac{1}{6 {\mathcal   H} \rho_{\rm t}'} \sum_{\rm i\,j} \rho_{\rm i}' \rho_{\rm j}' (c_{\rm s\,i}^2 
- c_{\rm s\,j}^2) S_{\rm i\,j},\qquad c_{\rm s}^2 = \frac{p_{\rm t}'}{\rho_{\rm t}'}.
\label{naddef}
\end{equation}
In Eq. (\ref{naddef}) $c_{\rm s\,i}^2$ and $c_{\rm s\,j}^2$ are the sound speeds of each 
(generic) pair of fluids of the mixture. According to Eq. (\ref{nad}), the 
fluctuation in the total pressure may arise either because of a 
inhomogeneity in the energy density or thanks to the chemical 
inhomogeneity of the plasma. Chemically inhomogeneous 
means, in this context, that the plasma is constituted 
by various  fluids with equations of state that are 
different and that entail  necessarily a spatial variation 
of the sound speed. As a consequence of Eqs. (\ref{nad}) and (\ref{naddef}), 
the evolution of $\zeta$ can be directly obtained from the evolution 
of $\delta_{\rm s}\rho_{\rm t}$ 
stemming from the first-order covariant conservation equation:
\begin{equation}
\delta_{\rm s} \rho_{\rm t}' + 3 {\mathcal   H} (\delta_{\rm s} \rho_{\rm t} + \delta_{\rm s} p_{\rm t}) - 3 \psi' (\rho_{\rm t} + p_{\rm t}) + (p_{\rm t} + \rho_{\rm t}) 
\theta_{\rm t} = \frac{\vec{E}\cdot \vec{J}}{ a^4}.
\end{equation}
Neglecting the Ohmic electric field we do obtain the wanted evolution of $\zeta$, i.e. 
\begin{equation}
\zeta' = - \frac{{\mathcal   H}}{p_{\rm t} +\rho_{\rm t}} \delta p_{\rm nad} + \frac{{\mathcal   H}}{p_{\rm t} + \rho_{\rm t}} \biggl( c_{\rm s}^2 - \frac{1}{3}\biggr) \delta\rho_{\rm B} - \frac{\theta_{\rm t}}{3}.
\label{zetaevol}
\end{equation}

By using Eqs. (\ref{zetaevol}) and (\ref{zetapar}) together 
with the explicit form of the evolution equations 
in the longitudinal gauge, the analytic form of the 
non-adiabatic contributions can be obtained. 
In the following, as an example, 
 the magnetized CDM-radiation mode and the magnetized neutrino entropy mode will be presented.
 For the case of the CDM-radiation mode the solution, in the limit 
$\tau <\tau_{1}$ and $k\tau < 1$ can be written as 
\begin{eqnarray}
&& \phi= \phi_{1} \biggl(\frac{\tau}{\tau_{1}}\biggr),\qquad  \psi= \psi_{1} \biggl(\frac{\tau}{\tau_{1}}\biggr),
\nonumber\\
&& \delta_{\gamma} = \delta_{\nu} = 4 \psi_{1} \biggl(\frac{\tau}{\tau_{1}}\biggr)- 
R_{\gamma}\Omega_{\rm B},
\nonumber\\
&& \delta_{\rm c} = - \biggl[ {\mathcal   S}_{*} + \frac{3}{4} R_{\gamma} \Omega_{\rm B}\biggr] + 3 \psi_{1} \biggl(\frac{\tau}{\tau_{1}}\biggr),\qquad \delta_{\rm b} = 3 \psi_{1} \biggl(\frac{\tau}{\tau_{1}}\biggr) - \frac{3}{4} R_{\gamma} \Omega_{\rm B},
\nonumber\\
&& \theta_{\rm c} = \frac{k^2 \tau_{1}}{3} \phi_{1} \biggl(\frac{\tau}{\tau_{1}}\biggr)^2 ,\qquad \theta_{\gamma{\rm b}} =\frac{k^2 \tau_1}{2} (\phi_1 + \psi_{1}) \biggl(\frac{\tau}{\tau_{1}}\biggr)^2  + \frac{k^2 \tau}{4} [ R_{\nu} \Omega_{\rm B} - 4 \sigma_{\rm B}],
\nonumber\\
&& \theta_{\nu} = \frac{k^2 \tau_1}{2} (\phi_1 + \psi_{1}) \biggl(\frac{\tau}{\tau_{1}}\biggr)^2  + \frac{k\tau}{4} \biggl(4 \frac{R_{\gamma}}{R_{\nu}} \sigma_{\rm B} - \Omega_{\rm B}\biggr),
\nonumber\\
&&  {\mathcal   F}_{\nu3} = \frac{8}{9} k\tau \biggl[ 4 \frac{R_{\gamma}}{R_{\nu}} \sigma_{\rm B} 
-\Omega_{\rm B}\biggr],\qquad \sigma_{\nu} = -\frac{R_{\gamma}}{R_{\nu}} \sigma_{\rm B} + 
\frac{k^2 \tau_{1}^2}{6 R_{\nu}} ( \psi_{1} -\phi_1) \biggl(\frac{\tau}{\tau_{1}}\biggr)^3,
\label{CDMNAD1}
\end{eqnarray}
where 
\begin{equation}
\psi_{1} = \frac{15 + 4 R_{\nu}}{8( 15 + 2 R_{\nu})}\biggl[ {\mathcal   S}_{*} + 
\frac{3}{4} R_{\gamma} \Omega_{\rm B} \biggr], \qquad
\phi_{1} = \frac{15 - 4 R_{\nu}}{8( 15 + 2 R_{\nu})}\biggl[ {\mathcal   S}_{*} + 
\frac{3}{4} R_{\gamma} \Omega_{\rm B} \biggr].
\label{CDMNAD2}
\end{equation}
In the case of this solution, the longitudinal fluctuations 
of the geometry are vanishing for $\tau < \tau_{1}$ and the relevant 
entropy fluctuations, i.e. ${\mathcal   S}_{{\rm c}\gamma}$ and ${\mathcal   S}_{{\rm c}\nu}$ are constant and proportional to ${\mathcal   S}_{*}$. 
In the limit $\Omega_{\rm B}\to 0$ and $\sigma_{\rm B}\to 0$ 
the solution (\ref{CDMNAD1}) reduces to the conventional CDM-radiation 
mode. While in the standard case 
the octupole moment of the (perturbed) neutrino phase-space 
distribution vanishes, in the magnetized case it is 
proportional to $k\tau$ and is negligible for length-scales 
larger than the Hubble radius prior to equality.
In analogy with Eq. (\ref{RSP}), the spectrum of entropy 
perturbations will be parametrized as 
\begin{equation}
\qquad \langle {\mathcal   S}_{*}(\vec{k}) {\mathcal  S}_{*}(\vec{p}) \rangle = \frac{2\pi^2 }{k^3} {\mathcal   P}_{\mathcal   S}(k) \delta^{(3)}(\vec{k} + \vec{p}),
 \qquad
{\mathcal  P}_{{\mathcal   S}}(k) = A_{{\mathcal   S}} \biggl(\frac{k}{k_{\rm p}}\biggr)^{n_{s} -1},
\label{SSP}
\end{equation}
where $k_{\rm p}$ is the same pivot scale introduced in Eq. (\ref{RSP}) while 
$n_{s}$ is the spectral index of the entropy fluctuation.
In the case of the magnetized neutrino entropy mode, 
the solution can instead be presented as:
\begin{eqnarray}
&& \phi= \phi_{0},\qquad \psi=-\frac{ \phi_{0}}{2}, \qquad \phi_{0} = \frac{8 R_{\gamma} R_{\nu}}{3 ( 4 R_{\nu} + 15)}\biggl[ \tilde{{\mathcal   S}}_{*} - \frac{3}{R_{\nu}} \sigma_{\rm B} + 
\frac{3}{4} \Omega_{\rm B}\biggr],
\nonumber\\
&& \delta_{\gamma} =  - 2 \phi_{0} + \frac{4}{3} \tilde{{\mathcal   S}}_{*} R_{\nu} - \Omega_{\rm B} R_{\gamma},\qquad  \delta_{\nu}=- 2 \phi_{0} - \frac{4}{3} \tilde{{\mathcal S}}_{*} R_{\gamma} - \Omega_{\rm B} R_{\gamma},
\nonumber\\
&&  \delta_{\rm b} = \delta_{\rm c} = -\frac{3}{2} \phi_{0} - \frac{3}{4} \Omega_{\rm B} R_{\gamma},
\nonumber\\
&& \theta_{\rm c} = \frac{k^2 \tau}{2} \phi_{0},\qquad\theta_{\gamma{\rm b}} = 
\frac{k^2 \tau}{4} \biggl[ 2 \phi_{0} - \frac{4}{3} \tilde{{\mathcal   S}}_{*} R_{\nu} + R_{\nu} \Omega_{\rm B} - 4 \sigma_{\rm B}\biggr],
\nonumber\\
&& \theta_{\nu} = \frac{k^2 \tau}{4} \biggl[ 2 \phi_0 - \frac{4}{3} \tilde{\mathcal   S}_{*}(R_{\gamma} +1)  + 
4 \frac{R_{\gamma}}{R_{\nu}} \sigma_{\rm B} - \Omega_{\rm B} R_{\gamma} \biggr]
\nonumber\\
&& \sigma_{\nu} = - \frac{R_{\gamma}}{R_{\nu}}  \sigma_{\rm B} - \frac{k^2 \tau^2 }{4 R_{\nu}} \phi_{0}.
\label{NUE1}
\end{eqnarray}
In the case of the neutrino-entropy mode ${\mathcal   S}_{\nu\gamma}\neq 0$ 
and it is proportional to a constant that has been denoted 
in Eq. (\ref{NUE1}) as $\tilde{{\mathcal   S}}_{*}(k)$; the spectrum of 
$\tilde{{\mathcal S}}_{*}(k)$ can be defined in full analogy 
with Eq. (\ref{SSP}).  An interesting feature 
of the magnetized neutrino-entropy mode (common to the 
conventional case when magnetic fields are absent) is that the 
curvature perturbations on comoving orthogonal hypersurfaces 
(i.e. ${\mathcal   R}$ defined in Eq. (\ref{calardef})) vanish for 
wavelengths larger than the Hubble radius before equality. This 
statement can be verified by using Eq. (\ref{NUE1}) into Eq. (\ref{calardef}) 
with the result that ${\mathcal   R} = - \psi - \phi/2 =0$ for $k\tau \ll 1$.

\renewcommand{\theequation}{3.\arabic{equation}}
\section{Higher-order tight coupling expansion}
\setcounter{equation}{0}

After having taken the derivative with respect to the conformal time $\tau$ of the first relation appearing in Eq. (\ref{gamma1}),
 Eq. (\ref{TC2}) can be used to eliminate the baryon-photon velocity; 
 the result will of this manipulation is:
\begin{equation}
\delta_{\gamma}'' + \frac{{\mathcal   H} R_{\rm b}}{R_{\rm b} + 1} \delta_{\gamma}' + 
\frac{k^2}{3 ( R_{\rm b} + 1)} \delta_{\gamma} = 4 \biggl[ \psi'' + 
\frac{{\mathcal   H} R_{\rm b}}{R_{\rm b} + 1} \psi' - \frac{k^2}{3} \phi\biggr] + 
\frac{k^2 }{3 ( R_{\rm b} + 1)} [ 4 \sigma_{\rm B} - \Omega_{\rm B}],
\label{TCd}
\end{equation}
where, according to Eq. (\ref{an1}), the Lorentz force has been traded for a 
combination of the magnetic energy density and of the normalized 
anisotropic stress. Equation (\ref{TCd}) stems directly from the 
zeroth order in the tight-coupling expansion. 

To discuss the polarization, we have to go (at least) to first-order in the tight 
coupling expansion. For this purpose, it is appropriate to introduce 
the evolution equations of the brightness perturbations of the $I$, $Q$ and $U$ Stokes parameters characterizing the radiation field. Since the Stokes parameters $Q$ and $U$ are not invariant under rotations about the axis 
of propagation the degree of polarization $P = (Q^2 + U^2)^{1/2}$ is customarily introduced \cite{TC3,CH}. The relevant  brightness 
perturbations will then be denoted as $\Delta_{\rm I}$, $\Delta_{\rm P}$. This description, reproduces, to zeroth order 
in the tight coupling expansion, the fluid equations that have been presented 
in Section 2 to set initial conditions prior to equality. For 
instance, the photon density contrast and the divergence of the photon 
peculiar velocity are related, respectively, to the 
monopole and to the dipole of the brightness perturbation of the intensity field, 
i.e. $\delta_{\gamma} = 4 \Delta_{{\rm I}0}$ 
and  $\theta_{\gamma} = 3 k \Delta_{{\rm I}1}$. 
The evolution equations of the brightness perturbations can then 
be written, within the conventions set by Eqs. (\ref{met1}) and 
(\ref{longfl}) as
\begin{eqnarray}
&& \Delta_{\rm I}' + (i k \mu + \kappa') \Delta_{\rm I} + i k \mu \phi= 
\psi' + \kappa' \biggl[ \Delta_{{\rm I}0} + \mu v_{\rm b} - 
\frac{1}{2} P_{2}(\mu) S_{\rm P}\biggr],
\label{DI}\\
&& \Delta_{\rm P}' + ( i k \mu + \kappa') \Delta_{\rm P} = \frac{\kappa'}{2} 
[ 1 - P_{2}(\mu)] S_{\rm P},
\label{DQ}\\
&& v_{\rm b}' + {\mathcal   H} v_{\rm b} + i k \phi + \frac{i k}{4 R_{\rm b}} [\Omega_{\rm B} - 4 \sigma_{\rm B}] + \frac{\kappa'}{R_{\rm b}} ( v_{\rm b} + 3 i \Delta_{{\rm I}1})=0.
\label{vb}
\end{eqnarray}
Equation (\ref{vb}) is nothing but the second relation obtained in  Eq. (\ref{b1})
having introduced the quantity $i k v_{\rm b} = \theta_{\rm b}$. The source 
terms appearing in Eqs. (\ref{DI}) and (\ref{DQ}) include a dependence 
on $P_{2}(\mu)= (3 \mu^2 -1)/2$ ( $P_{\ell}(\mu)$ denotes, in this 
framework, the $\ell$-th Legendre polynomial);
 ; $\mu = \hat{k} \cdot\hat{n}$ is 
simply
the projection of the Fourier wave-number on the direction of the photon momentum. In Eqs. (\ref{DI}) and (\ref{DQ}) the source 
term $S_{\rm P}$ is defined as
\begin{equation}
S_{\rm P}(k,\tau) = \Delta_{{\rm I}2}(k,\tau) + \Delta_{{\rm P}0}(k,\tau) + \Delta_{{\rm P}2}(k,\tau).
\end{equation}
The evolution of the longitudinal fluctuations is dictated by Eq. (\ref{anis2})
and by Eqs. (\ref{ham0})--(\ref{mom0}). According to the definitions used to derive 
Eqs. (\ref{DI}), (\ref{DQ}) and (\ref{vb}),  the 
$\ell$-th multipole of the brightness perturbations is defined as 
\begin{equation}
\int_{-1}^{1} P_{\ell}(\mu) \Delta_{\rm I}(k,\mu,\tau) d\mu = 
2 (-i)^{\ell} \Delta_{{\rm I}\ell}(k,\tau),
\qquad \int_{-1}^{1} P_{\ell}(\mu) \Delta_{\rm P}(k,\mu,\tau) d\mu = 
2 (-i)^{\ell} \Delta_{{\rm P}\ell}(k,\tau).
\label{defmult}
\end{equation}
Equations (\ref{DI}) and (\ref{DQ})  con now be expanded in powers 
of $\tau_{\rm c} = |1/\kappa'|$. In particular we can write
\begin{eqnarray}
&&\Delta_{{\rm I}\ell}(k,\tau) = \overline{\Delta}_{{\rm I}\ell}(k,\tau) + \tau_{\rm c} \delta_{{\rm I}\ell}(k,\tau),
\qquad \Delta_{{\rm P}\ell}(k,\tau) =  
\overline{\Delta}_{{\rm P}\ell}(k,\tau) + \tau_{\rm c} \delta_{{\rm P}\ell}(k,\tau),
\nonumber\\ 
&&v_{\rm b} = \overline{v}_{\rm b}(k,\tau) + \tau_{\rm c} \delta_{v_{\rm b}}(k,\tau).
\label{exp}
\end{eqnarray}
Equations (\ref{DI}) and (\ref{DQ}) 
will be preliminarily phrased in terms 
of a Boltzmann hierarchy coupling together the different multipoles 
of the brightness perturbations.  Then the obtained equations 
will be expanded according to (\ref{exp}). The result of the first step of this 
procedure is 
\begin{eqnarray}
&& \Delta_{{\rm I}0}' + k \Delta_{{\rm I}1} = \psi', 
\label{mon}\\
&& \Delta_{{\rm I}1}' + \frac{2}{3} k \Delta_{{\rm I}2} - \frac{k}{3} \Delta_{{\rm I}0} = 
\frac{k}{3} \phi  - \kappa' \biggl[ \Delta_{{\rm I}1} + \frac{v_{\rm b}}{3 i}\biggr],
\label{dip}\\
&& \Delta_{{\rm I}2}' + \frac{3}{5} k \Delta_{{\rm I}3} - \frac{2}{5} k \Delta_{{\rm I}1} = 
- \frac{\kappa'}{10}[ 9 \Delta_{{\rm I}2}  - (\Delta_{{\rm P}0} + \Delta_{{\rm P}2})],
\label{quadr}\\
&& \Delta_{{\rm I}\ell}' + \kappa' \Delta_{{\rm I}\ell} 
= \frac{k}{2 \ell + 1} [ \ell \Delta_{{\rm I}(\ell-1)} - (\ell + 1) \Delta_{{\rm I}(\ell + 1)}],\qquad \ell > 2.
\label{oct}
\end{eqnarray}
Equations (\ref{mon}), (\ref{dip}) and (\ref{quadr}) are obtained, respectively,  by multiplying each
side of Eq. (\ref{DI}) by $P_{0}(\mu)$, $P_{1}(\mu)$ and $P_{2}(\mu)$ and by integrating 
over $\mu$ with the help of  Eq. (\ref{defmult}).
The same calculation can be repeated in the case of Eq. (\ref{DQ}); the result is:
\begin{eqnarray}
&& \Delta_{{\rm P}0}' + k \Delta_{{\rm P}1} = \frac{\kappa'}{2}[ \Delta_{{\rm P}2} + 
\Delta_{{\rm I}2} - \Delta_{{\rm P}0}],
\label{monQ}\\
&& \Delta_{{\rm P}1}'  + \frac{2}{3} k \Delta_{{\rm P}2} - \frac{k}{3} \Delta_{{\rm P}0} = - \kappa' \Delta_{{\rm P}1},
\label{dipQ}\\
&& \Delta_{{\rm P}2}' + \frac{3}{5} k \Delta_{{\rm P}3} - \frac{2}{5} k 
\Delta_{{\rm P}1}  = - \frac{\kappa'}{10}[ 9 \Delta_{{\rm P}2} - (\Delta_{{\rm P}0} +\Delta_{{\rm I}2})],
\label{quadrQ}\\
&& \Delta_{{\rm P}\ell}' + \kappa' \Delta_{{\rm P}\ell} 
= \frac{k}{2 \ell + 1} [ \ell \Delta_{{\rm P}(\ell-1)} - (\ell + 1) \Delta_{{\rm P}(\ell + 1)}],\qquad \ell > 2.
\label{octQ}
\end{eqnarray}
Having derived Eqs. (\ref{mon})--(\ref{oct}) and Eqs. (\ref{monQ})--(\ref{octQ}), the system can now be expanded  
to the wanted order in $\tau_{\rm c}$.  As anticipated, 
to zeroth order in the tight coupling expansion Eq. (\ref{TCd}) can be 
reobtained. In fact, to lowest order in $\tau_{\rm c}$,
\begin{equation}
\overline{v}_{\rm b}(k,\tau) = - 3 i \overline{\Delta}_{{\rm I}1}(k,\tau),\qquad \overline{\Delta}_{{\rm I}0}' + 
k \overline{\Delta}_{{\rm I}1} = \psi',
\label{zerothTC1}
\end{equation}
while, from Eqs. (\ref{monQ})--(\ref{octQ}),  
\begin{equation}
\overline{\Delta}_{{\rm P}0}(k,\tau) =0,\,\,\,\, \ell \geq 0,\qquad 
 \overline{\Delta}_{{\rm I}\ell}(k,\tau) =0, \,\,\,\,
\ell \geq 2.
\label{zerothTC2}
\end{equation}
Summing up Eq. (\ref{mon}) (multiplied by $3 i$) with Eq. (\ref{vb}) (multiplied 
by $R_{\rm b}$) and  eliminating, from the obtained relation, 
$\overline{\Delta}_{{\rm I}1}$ in favour of $\overline{\Delta}_{{\rm I}0}'$  (see Eq. 
(\ref{zerothTC1})) the decoupled expression for the evolution of the zeroth-order monopole becomes, as expected,
\begin{equation}
\overline{\Delta}_{{\rm I}0}'' + \frac{{\mathcal   H} R_{\rm b}}{R_{\rm b} + 1} 
\overline{\Delta}_{{\rm I}0}' + \frac{k^2}{3 ( R_{\rm b} + 1)} \overline{\Delta}_{{\rm I}0} = 
\biggl( \psi'' + \frac{{\mathcal   H} R_{\rm b}}{1 + R_{\rm b}} \psi' - \frac{k^2}{3} \phi \biggr) + 
\frac{k^2 ( 4 \sigma_{\rm B} - \Omega_{\rm B})}{3 (R_{\rm b} + 1)},
\label{mon0}
\end{equation}
which is the same of Eq. (\ref{TCd}) if we recall that $4 \overline{\Delta}_{{\rm I}0} = \delta_{\gamma}$. The presence of the magnetic field modifies 
the evolution of the zeroth order monopole and, consequently, according 
to Eq. (\ref{zerothTC1}), the zeroth order dipole.  Such a  
modification also implies an effect on  the polarization. 
In fact, to first-order in the tight coupling expansion, $\Delta_{{\rm P}0} \neq 0$ and it 
is proportional to the zeroth order dipole.  Recalling the notations 
of Eq. (\ref{exp}) and expanding Eqs. (\ref{mon})--(\ref{oct}) and 
(\ref{monQ})--(\ref{octQ})
to first order in $\tau_{\rm c}$ it can be verified that 
\begin{equation}
\delta_{{\rm P}0}(k,\tau) = \frac{5}{4} \delta_{{\rm I}2}(k,\tau),\qquad \delta_{{\rm P}2}(k,\tau) = \frac{1}{4} 
\delta_{{\rm I}2}(k,\tau),\qquad \delta_{{\rm I}2}(k,\tau) = \frac{8}{15} k 
\overline{\Delta}_{\rm I1}(k,\tau),
\label{FTC1}
\end{equation}
which also implies
\begin{equation}
\Delta_{{\rm I}2}(k,\tau)\simeq \tau_{\rm c} \delta_{{\rm I} 2}(k,\tau) \simeq 
\frac{8}{15} \,\, k\tau_{\rm c} \overline{\Delta}_{{\rm I}1}(k,\tau).
\end{equation}
Since $\tau_{\rm c}$ grows very fast during decoupling, to have a better 
quantitative estimate of the effect the procedure outlined in \cite{TC3}
can be followed. 
By taking the first derivative of the definition of $S_{\rm P}$ and by using 
Eqs. (\ref{oct}) and (\ref{octQ}) we can get, directly the following 
evolution equation for $S_{\rm P}$ 
\begin{equation}
S_{\rm P}'  +  \frac{3}{10} \kappa' S_{\rm P} = \frac{2}{5} k \overline{\Delta}_{{\rm I}1}. 
\label{eff}
\end{equation}
Equation (\ref{eff}) can be used to estimate the source term in Eq. (\ref{DQ}) 
and, therefore, the polarization can be determined as 
\begin{equation}
\Delta_{{\rm P}}(k,\tau_{\rm dec}) = - \frac{3}{10}  ( 1 - \mu^2) \overline{\Delta}_{{\rm I}1}(\tau_{\rm dec})
k \sigma_{\rm dec} {\mathcal   I}_{1},
\label{eff2}
\end{equation}
where $\sigma_{\rm dec} \simeq 70$--$80\,\,{\rm Mpc}$ is the width of the visibility function $\kappa' e^{-\kappa}$ which has a Gaussian form\cite{JO}.
The quantity ${\mathcal I}_{1}$ appearing in Eq. (\ref{eff2}) is 
\begin{equation}
{\mathcal   I}_{1} = \int_{0}^{\infty} e^{-\frac{7}{10} \kappa} d \kappa \int_{1}^{\infty} 
\frac{d x}{x} e^{- \frac{3}{10} \kappa x} \simeq 1.719.
\label{eff3}
\end{equation}

Let us now perform an oversimplified estimate. Let us ignore neutrinos and magnetic fields and let us compute out of Eq. (\ref{DI}) the ordinary Sachs-Wolfe contribution 
assuming a constant adiabatic mode that propagates smoothly through the matter-radiation transition. The constant mode will be $\psi_{\rm m} \simeq - (3/5) {\mathcal   R}_{*}$ 
where ${\mathcal   R}_{*}$ is the constant mode of curvature fluctuations.  Then, 
from Eqs. (\ref{mon0}) and (\ref{eff2})--(\ref{eff3})  
\begin{equation}
\Delta_{\rm I}(k,\tau_{\rm dec}) \simeq \frac{\psi_{\rm m}}{3} \cos{(k c_{\rm sb}\tau_{\rm dec})},\qquad 
\Delta_{{\rm P}}(k,\tau_{\rm dec}) = -0.17 (1- \mu^2)( \sigma_{\rm dec} k c_{\rm sb}) \psi_{\rm m} \sin{(k c_{\rm sb} \tau_{\rm dec})}.
\end{equation}
which implies that the cross-correlation between temperature and polarization oscillates as
 $\sin{( 2 k c_{\rm sb} \tau_{\rm dec})}$.

It is finally interesting to compute the expression of the characteristic damping scale of the fluctuations arising from the dispersion relations. From 
Eq. (\ref{vb}), by taking the Laplace transform, i.e. $v_{\rm b}(k,\tau) = e^{i \int \omega(\tau') d\tau'} v_{\rm b}(k,\omega)$ (and similarly for the brightness perturbations), we will have that the following 
relation holds to second order in the tight-coupling expansion
\begin{equation}
v_{\rm b}(k,\omega) = - 3 \, i\, \Delta_{{\rm I}1}(k,\omega) \biggl[ 1 - i \frac{\omega\,R_{\rm b}}{|\kappa'|} - \frac{\omega^2 R_{\rm b}^2}{|\kappa'|^2}\biggr]
+ {\cal O} (\phi) + {\cal O}( \Omega_{\rm B}, \sigma_{\rm B}).
\label{vb2}
\end{equation}
In the standard derivation of the dispersion relations the metric fluctuations are 
neglected. This procedure is also justified in the present case 
provided the spectrum of the magnetic energy density is not too steep 
in frequency. In practice this is the case for $\varepsilon$ sufficiently 
smaller than $1$. In this case also the magnetic contribution can be neglected.
In this situation, we can Laplace transform also the other equations 
(i.e. for the monopole, dipole and quadrupole). Then the first-order 
tight coupling expressions can be used into  Eqs. (\ref{dip}) 
and (\ref{vb2}) to eliminate $v_{\rm b}$. The resulting 
expression will then be 
\begin{equation}
- \omega^2 ( R_{\rm b} + 1) + \frac{k^2}{3} + i\frac{\omega}{|\kappa'|} 
\biggl[ \frac{16}{45} k^2 + \omega^2 R_{\rm b}^2\biggr] =0,
\end{equation}
which implies that $\omega = \omega_{1} + i \omega_{2}$ where 
\begin{equation}
\omega_{1} = \frac{k }{3( 1 + R_{\rm b})},\,\,\,\, \omega_{2} = \frac{k^2}{k_{\rm d}}, \,\,\,\, k_{\rm d}^{-2} = \int_{0}^{\tau} \frac{d \tilde{\tau}}{2 \kappa'(\tilde{\tau})} \biggl[ \frac{16}{45(R_{\rm b} (\tilde{\tau})+1} + \frac{R_{\rm b}^2(\tilde{\tau})}{3 ( 1 + R_{\rm b}(\tilde{\tau}))^2} \biggr].
\end{equation}
In practice, for the typical set of parameters introduced around Eq. (\ref{par}) 
$k_{\rm d}^{-1} \simeq 3 \,\, {\mathrm Mpc}$, which sets, grossly speaking, 
the typical scale of diffusion damping.  
  
\renewcommand{\theequation}{4.\arabic{equation}}
\section{Numerical analysis}
\setcounter{equation}{0}
The evolution equations in the tight coupling approximation will now be integrated 
numerically. 
The normalization of the numerical calculation is enforced by evaluating, 
analytically, the Sachs-Wolfe plateau and by deducing, for a given set 
of spectral indices of curvature and entropy perturbations, the 
amplitude of the power spectra at the pivot scale.  Here is an 
example of this strategy.
The Sachs-Wolfe (SW) plateau can  be estimated analytically 
from the evolution equation of ${\mathcal   R}$ (or $\zeta$) by using the technique of the transfer 
matrix appropriately generalized to the case where, on top of the adiabatic and non-adiabatic contributions the 
magnetic fields are consistently taken into account. The 
result of this calculation can be expressed,  for $\ell < 30$, as \cite{mg1}
\begin{eqnarray}
&&C_{\ell} = \biggl[ \frac{{\mathcal    A}_{\mathcal   R}}{25} \,{\mathcal   Z}_{1}(n_{r},\ell)  +
\frac{9}{100} \, R_{\gamma}^2  \overline{\Omega}^{2}_{{\rm B}\,L} {\mathcal   Z}_{2}(\epsilon,\ell) - 
\frac{4}{25} \sqrt{{\mathcal   A}_{\mathcal   R} {\mathcal  A}_{{\mathcal   S}}} {\mathcal   Z}_{1}(n_{r s},\ell) \cos{\gamma_{r s}}
\nonumber\\
&& +\frac{4}{25} \,{\mathcal   A}_{{\mathcal  S}} \,
{\mathcal   Z}_{1}(n_{s},\ell)- 
\frac{3}{25} \sqrt{ {\mathcal   A}_{\mathcal   R}} \, R_{\gamma} \,\overline{\Omega}_{{\rm B}\, L}\,{\mathcal   Z}_{3} (n_{r},\varepsilon, \ell) \cos{\gamma_{br}} 
\nonumber\\
&& + \frac{6}{25} \sqrt{ {\mathcal  A}_{{\mathcal  S}}}\,R_{\gamma} \overline{\Omega}_{{\rm B}\,L}\, {\mathcal   Z}_{3}(n_{s},\varepsilon, \ell)\cos{\gamma_{b s}} \biggr],
\label{SWP}
\end{eqnarray}
where the functions ${\mathcal   Z}_{1}$, ${\mathcal   Z}_{2}$ and ${\mathcal   Z}_{3}$ 
\begin{eqnarray}
{\mathcal  Z}_{1}(n,\ell) &=& \frac{\pi^2}{4} \biggl(\frac{k_0}{k_{\rm p}}\biggl)^{n-1} 2^{n} \frac{\Gamma( 3 - n) \Gamma\biggl(\ell + 
\frac{ n -1}{2}\biggr)}{\Gamma^2\biggl( 2 - \frac{n}{2}\biggr) \Gamma\biggl( \ell + \frac{5}{2} - \frac{n}{2}\biggr)},
\label{Z1}\\
{\mathcal  Z}_{2}(\varepsilon,\ell) &=& \frac{\pi^2}{2} 2^{2\varepsilon} {\mathcal  F}(\varepsilon) \biggl( \frac{k_{0}}{k_{L}}\biggr)^{ 2 \varepsilon} \frac{ \Gamma( 2 - 2 \varepsilon) 
\Gamma(\ell + \varepsilon)}{\Gamma^2\biggl(\frac{3}{2} - \varepsilon\biggr) \Gamma(\ell + 2 -\varepsilon)},
\label{Z2}\\
{\mathcal  Z}_{3}(n,\varepsilon,\ell) &=&\frac{\pi^2}{4} 2^{\varepsilon} 2^{\frac{n +1}{2}} \,\sqrt{{\mathcal  F}(\varepsilon)}\, \biggl(\frac{k_{0}}{k_{L}}\biggr)^{\varepsilon} \biggl(\frac{k_{0}}{k_{\rm p}}\biggr)^{\frac{n + 1}{2}} \frac{ \Gamma\biggl(\frac{5}{2} - \varepsilon - \frac{n}{2}\biggr) \Gamma\biggl( \ell + 
\frac{\varepsilon}{2} + \frac{n}{4} - \frac{1}{4}\biggr)}{\Gamma^2\biggl(\frac{7}{4} - \frac{{\varepsilon}}{2} - \frac{n}{4}\biggr)
\Gamma\biggl( \frac{9}{4} + \ell - \frac{\varepsilon}{2} - \frac{n}{4} \biggr)}.
\label{Z3}
\end{eqnarray}
We recall the meaning of the various spectral indices appearing in Eqs. (\ref{Z1}), (\ref{Z2}) and (\ref{Z3}).
The spectral index $\varepsilon$ has been introduced in Eq. (\ref{MPS}) (see also Eq. (\ref{POM}) for the 
magnetic energy spectrum); $k_{0} = \tau_{0}^{-1}$ where 
$\tau_{0}$ is the present observation time. In Eqs. (\ref{Z1}), (\ref{Z2}) and (\ref{Z3}), $n$ stands either for
 $n_{r}$ (i.e. the spectral index of the adiabatic mode defined in Eq. (\ref{RSP})) or for 
$n_{s}$ (i.e.  the spectral index of the entropic mode defined in Eq. (\ref{SSP})) or for $n_{rs}$ (i.e. 
the spectral index of the cross-correlation). In principle 
one could also introduce a different spectral index for each cross-correlation. However, following, for 
instance \cite{h3}  we will assume that the spectrum of the cross-correlation is solely determined 
in terms of the spectrum of the autocorrelations, i.e., for instance, $n_{rs} = (n_{r} + n_{s})/2$.
In Eq. (\ref{SWP}) $\gamma_{rs}$, $\gamma_{br}$ and $\gamma_{s b}$ are the correlation angles. 
From the request that the SW plateau is dominated by the adiabatic mode it is possible 
to constrain $\overline{\Omega}_{{\rm B} L}$ and, consequently, $B_{L}$ 
(see, indeed, Eq. (\ref{OMC2})).
The nature of the bound, however, depends on the parameters that enter Eq. (\ref{SWP}).
In the absence of magnetic and non-adiabatic contributions and for Eqs. (\ref{SWP}) 
and Eq. (\ref{Z1}) imply that for $n_{r}=1$ (Harrison-Zeldovich spectrum) $ \ell (\ell +1) C_{\ell}/2\pi = 
{\mathcal  A}_{{\mathcal R}}/25$ and WMAP data \cite{F5} would demand that 
${\mathcal  A}_{\mathcal R}= 2.65 \times 10^{-9}$. 

\begin{figure}
\begin{center}
\begin{tabular}{|c|c|}
      \hline
      \hbox{\epsfxsize = 7.6 cm  \epsffile{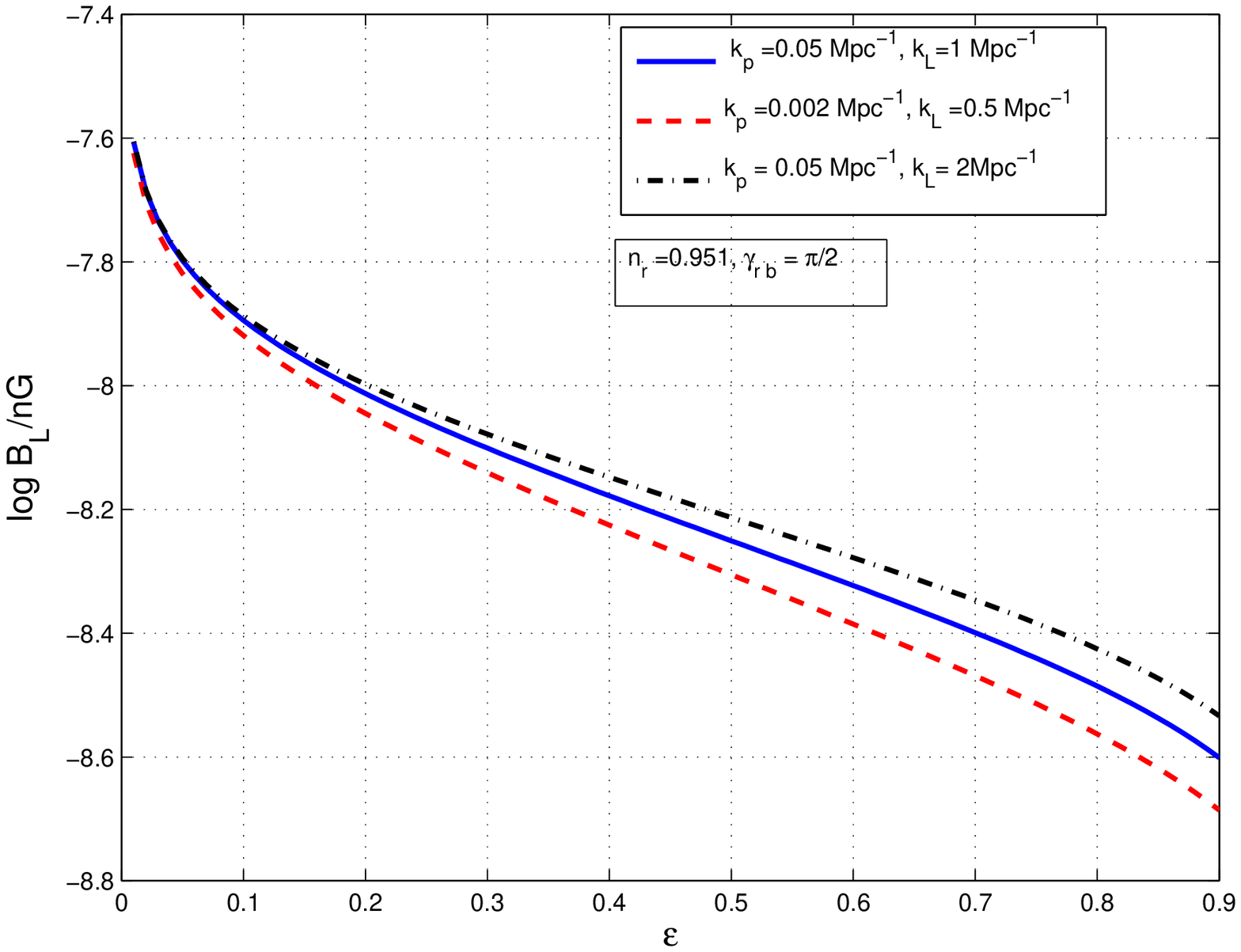}} &
      \hbox{\epsfxsize = 7.6 cm  \epsffile{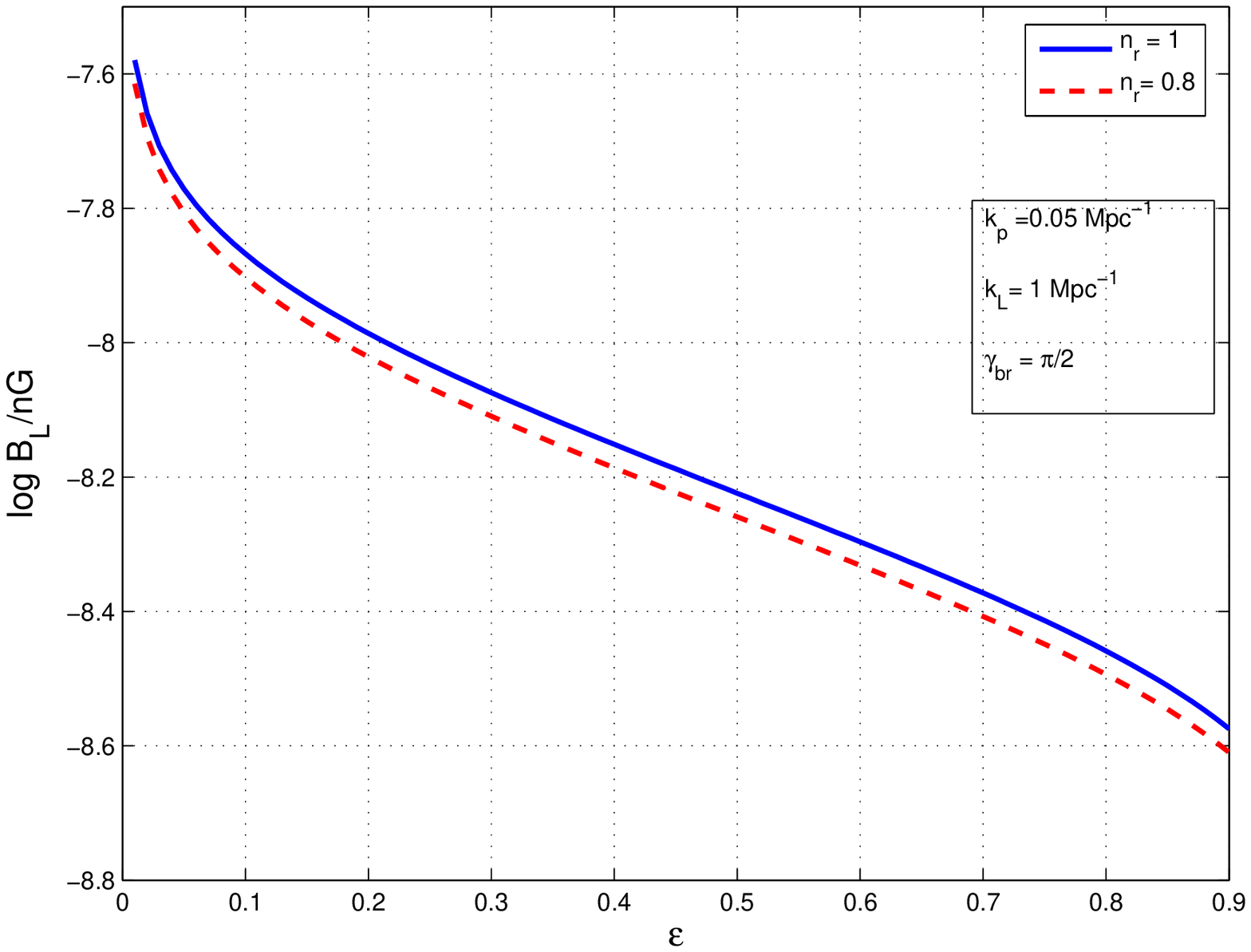}}\\
      \hline
\end{tabular}
\end{center}
\caption[a]{Bounds on the protogalactic field intensity as a function of the magnetic spectral index $\varepsilon$ for 
different values of the parameters defining the adiabatic contribution to the SW plateau.}
\label{F1}
\end{figure}

If the SW plateau is determined by an adiabatic component supplemented 
by a (subleading) non-adiabatic contribution both correlated with the magnetic field intensity the 
obtainable bound may not be so constraining 
(even well above the nG range) due to the proliferation of parameters. 
A possible strategy is therefore to fix the parameters of the adiabatic mode 
to the values determined by WMAP-3 and then explore the effect of a magnetized contribution 
which is not correlated with the adiabatic mode. This implies, in Eq. (\ref{SWP}) that ${\mathcal A}_{{\mathcal S}}=0$
and $\gamma_{\rm br} = \pi/2$. 
Under this assumption, in  Fig. \ref{F1}  the bounds on $B_{L}$ are illustrated. The nature of the 
constraint depends, in this case, both on the amplitude of the protogalactic
field  (at the present epoch and smoothed over a typical comoving scale 
$L = 2\pi/k_{L}$) and upon its spectral slope, i.e. $\varepsilon$. In the case $\varepsilon < 0.5$ the magnetic energy 
spectrum is nearly scale-invariant. In this case, diffusivity effects are negligible (see, for instance, \cite{r1,r2}). 
As already discussed, if $\varepsilon \gg 1$ the diffusivity effects (both thermal and magnetic)  dominate the mode-coupling integral that lead to the magnetic energy 
spectrum \cite{r1,r2}.
In Fig. \ref{F1} (plot at the left) the magnetic field intensity should be below the different curves if the 
adiabatic contribution dominates the SW plateau. Different choices of the pivot scale $k_{\rm p}$ and of the smoothing scale $k_{L}$, are also illustrated. 
In Fig. \ref{F1} (plot at the left) the scalar spectral index is fixed to $n_{r} = 0.951$ \cite{wmap3}.
In the plot at the right the two curves corresponding, respectively, to $n_{r}=0.8$ and $n_{r} =1$ are reported.
If $\varepsilon <0.2$ the bounds are comparatively less restrictive than in the case $\varepsilon \simeq 0.9$. 
The cause of this occurrence is that we are here just looking at the largest wavelengths of the problem. As it will become 
clear in a moment, intermediate scales will be  more sensitive to the presence of fully inhomogeneous magnetic fields. 

According to Fig. \ref{F1} for a given value of the magnetic spectral index and of the scalar spectral index 
the amplitude of the magnetic field has to be sufficiently small not to affect the dominant adiabatic nature of the SW plateau. 
Therefore Fig. \ref{F1} (as well as other similar plots) can be  used 
to normalize the numerical calculations for the power spectra of the brightness perturbations, i.e. 
\begin{equation}
\frac{k^{3}}{2\pi^2} |\Delta_{\rm I}(k,\tau)|^2,\qquad \frac{k^{3}}{2\pi^2} |\Delta_{\rm P}(k,\tau)|^2,\qquad
\frac{k^{3}}{2\pi^2} |\Delta_{\rm I}(k,\tau) \Delta_{\rm P}(k,\tau)|.
\label{defPS}
\end{equation}
Let us then assume, for consistency with the cases reported in Fig. \ref{F1}, that we are dealing with the situation 
where the magnetic field is not correlated with the adiabatic mode. It is then possible to choose 
a definite value of the magnetic spectral index (for instance $\epsilon = 0.1$) and a definite value 
of the adiabatic spectral index, i.e. $n_{r}$ (for instance $n_{r} =0.951$, in agreement 
with \cite{wmap3}).  By using the SW plateau the normalization can be chosen in such a way the the adiabatic mode dominates 
over the magnetic contribution. In the mentioned case, Fig. \ref{F1} implies $B_{L} < 1.14 \times 10^{-8}\,\,{\rm G}$ for a pivot 
scale $k_{p} = 0.002\, {\mathrm Mpc}^{-1}$. Since the relative weight of the power spectra given in Eqs. (\ref{RSP}) 
and (\ref{POM}) is fixed, it is now possible to set initial conditions for the adiabatic mode according to Eqs. 
(\ref{DCad})--(\ref{longad}) and (\ref{longad2}) deep in the radiation-dominated phase.  The initial time 
of integration will be chosen as $\tau_{\rm i} = 10^{-6} \tau_{1}$ in the notations discussed in Eq. (\ref{SF}).
According to Eq. (\ref{SF1}), this choice implies that $\tau_{\rm i} \ll \tau_{\rm eq}$.

The power spectra of the brightness perturbations, i.e. 
Eq. (\ref{defPS}), can be then computed by numerical integration. Clearly the calculation will depend 
upon the values of $\omega_{\rm m}$, $\omega_{\rm b}$, $\omega_{\rm c}$ and $R_{\nu}$. We will simply 
fix these parameters to their fiducial values reported in Eqs. (\ref{par}) (see also (\ref{RB})) and we will take $N_{\nu} = 3$ 
in Eq. (\ref{RNU}) determining, in this way the fractional contribution of the neutrinos to the radiation plasma.
\begin{figure}
\begin{center}
\begin{tabular}{|c|c|}
      \hline
      \hbox{\epsfxsize = 7.5 cm  \epsffile{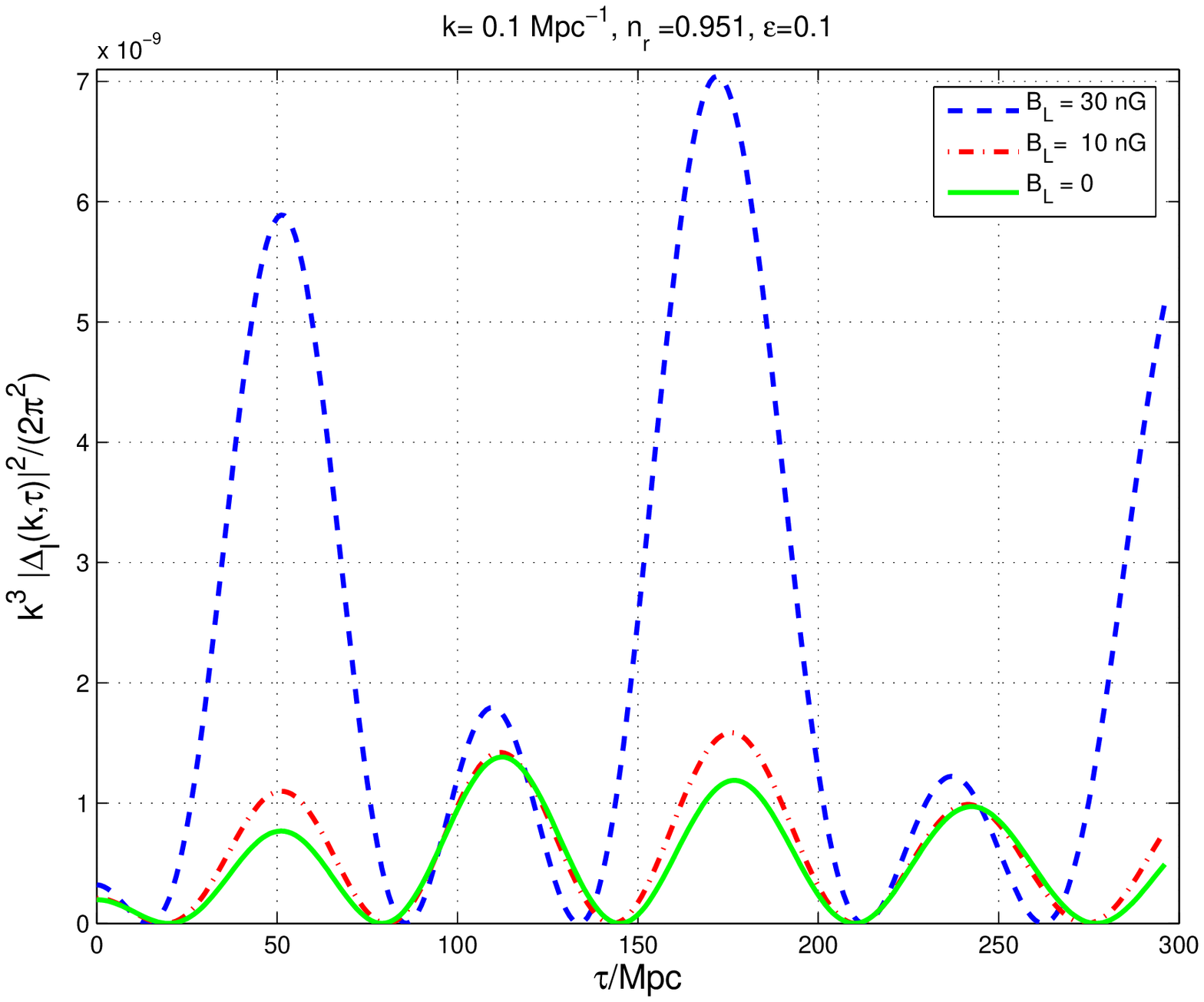}} &
      \hbox{\epsfxsize = 7.5 cm  \epsffile{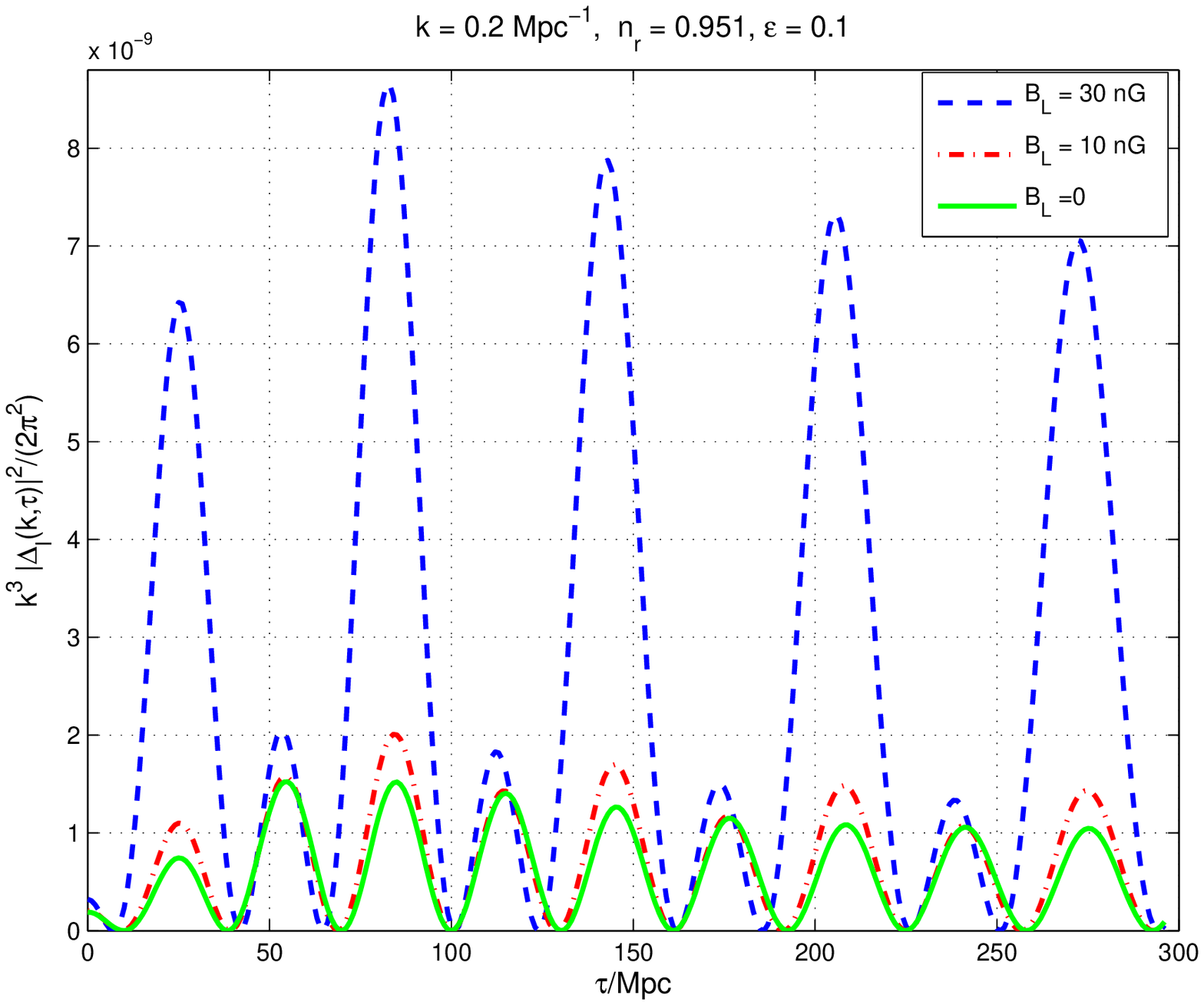}} \\
      \hline
      \hbox{\epsfxsize = 7.5 cm  \epsffile{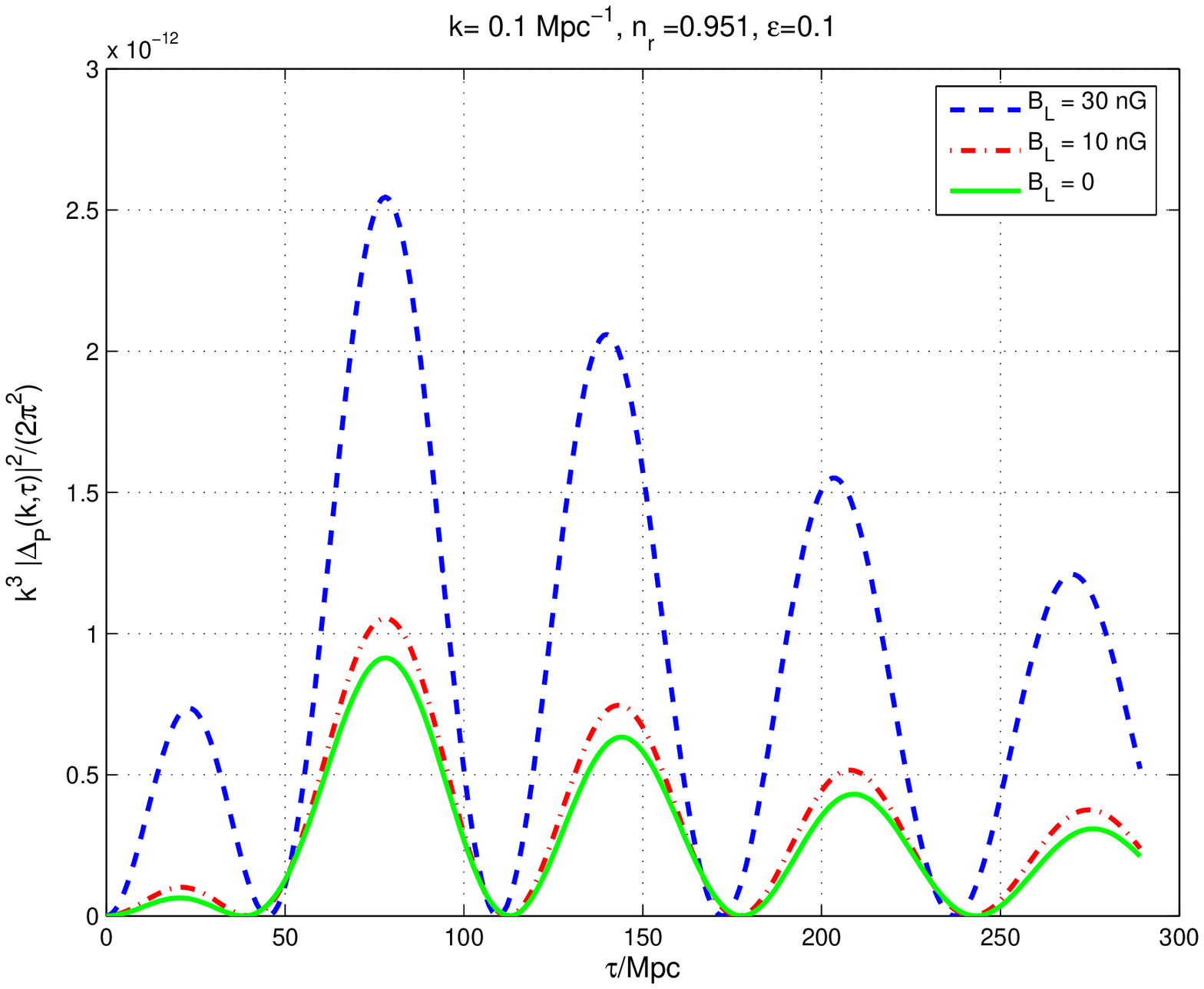}}  &
      \hbox{\epsfxsize = 7.5 cm  \epsffile{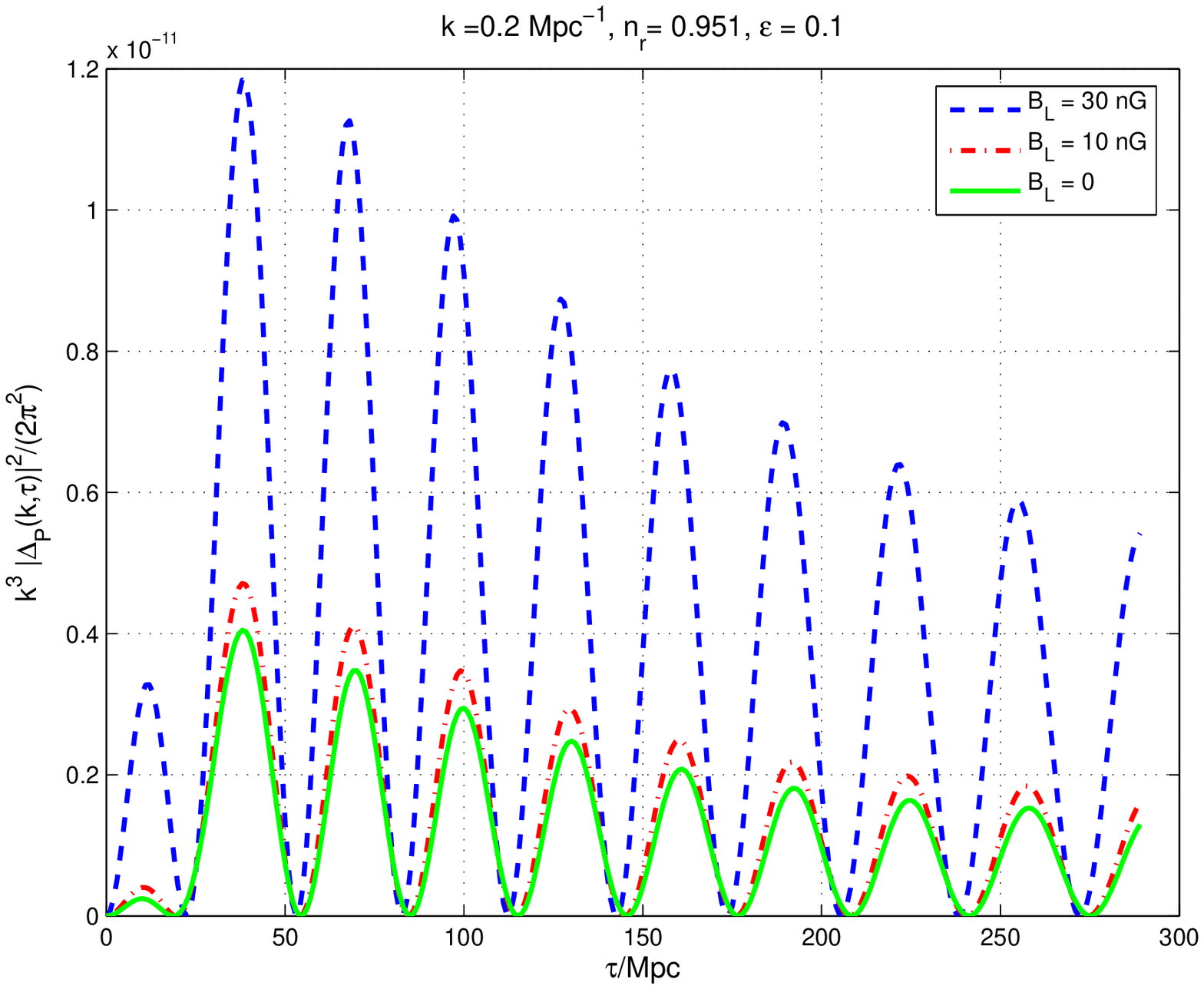}}\\
      \hline
      \hbox{\epsfxsize = 7.5 cm  \epsffile{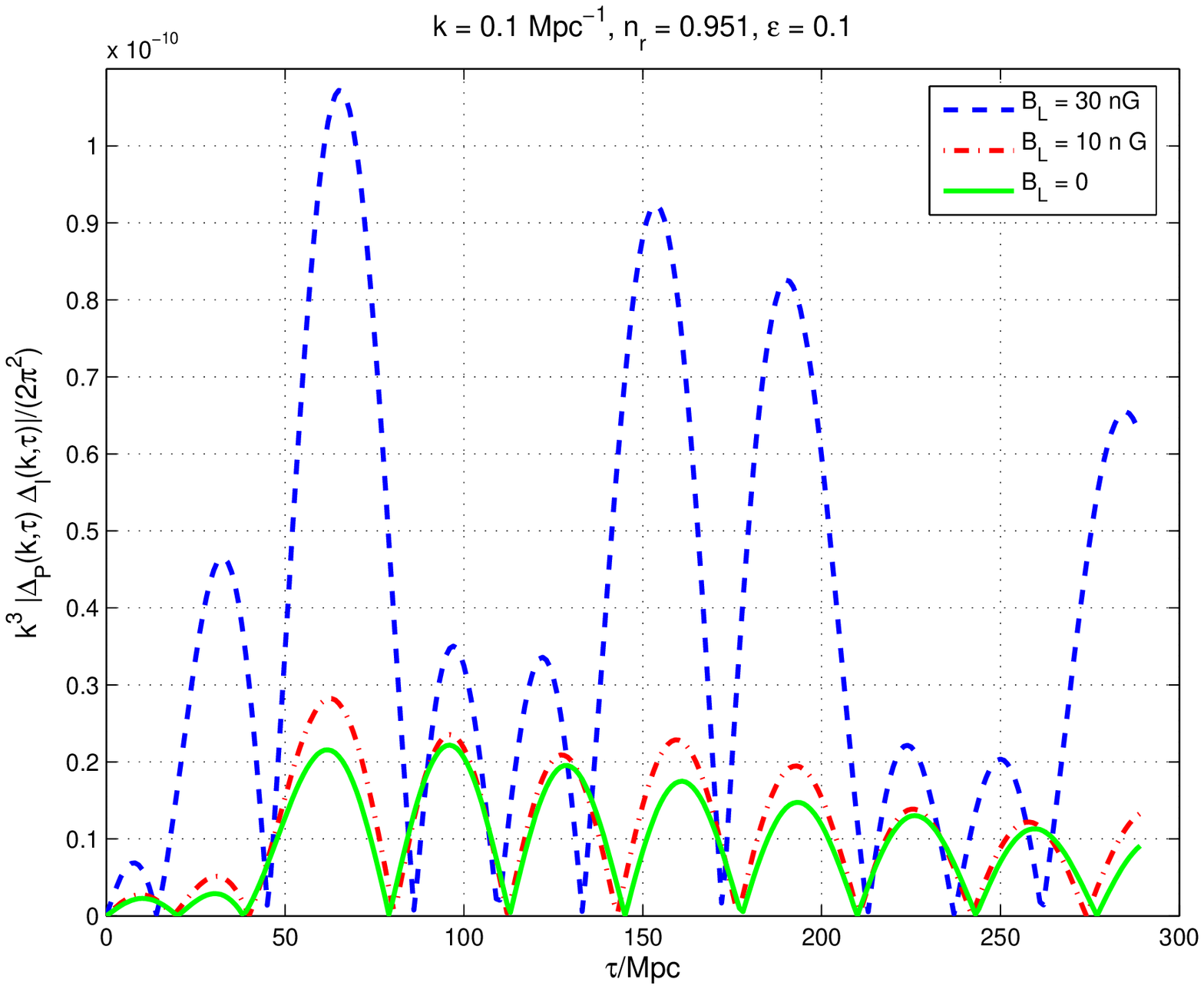}}  &
      \hbox{\epsfxsize = 7.5 cm  \epsffile{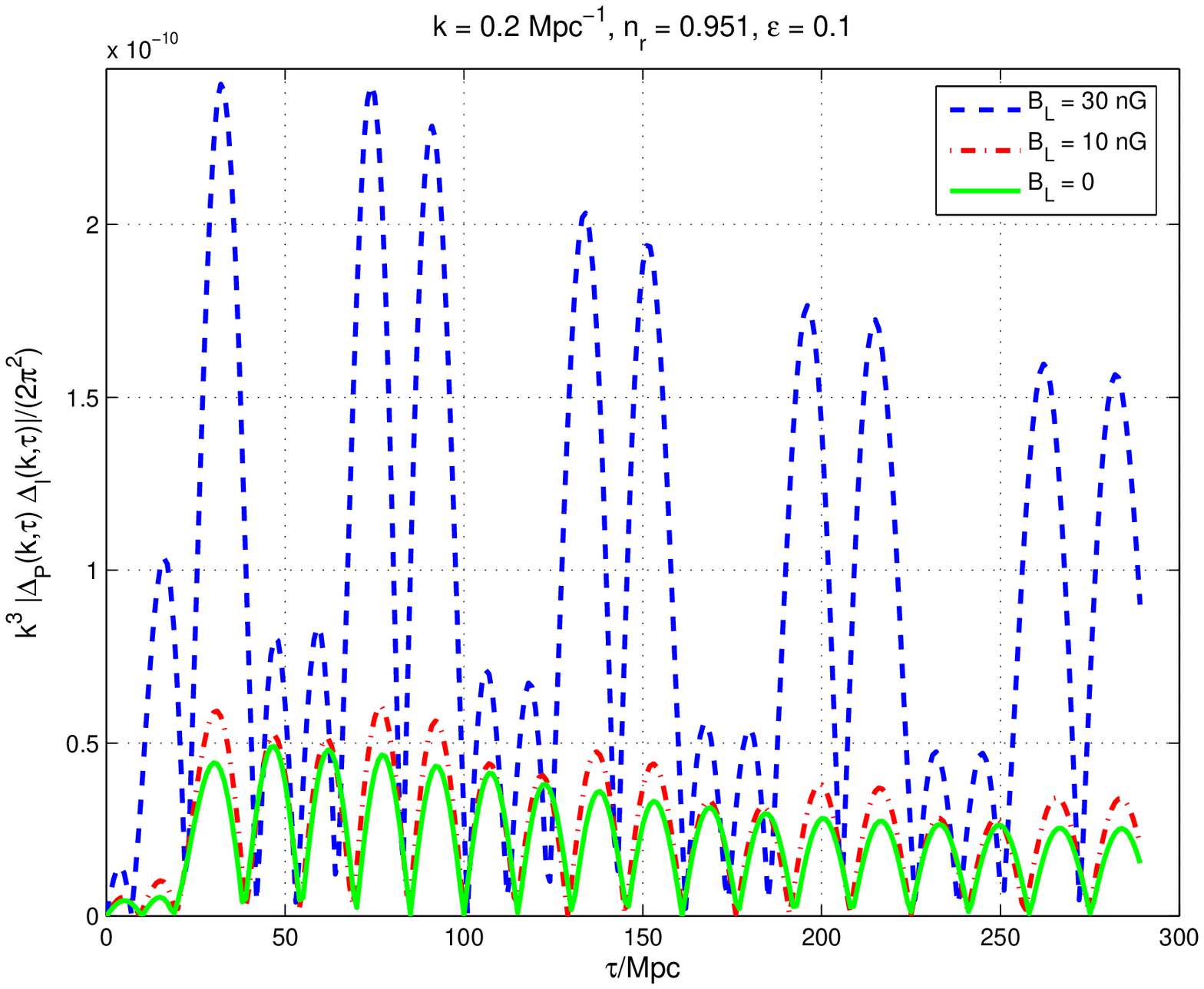}}\\
      \hline
\end{tabular}
\end{center}
\caption[a]{The power spectra of the brightness perturbations for two typical wave-numbers. The values 
of the parameters are specified in the legends. The pivot scale is $k_{\rm p} = 0.002\,\, {\mathrm Mpc}^{-1}$ and the 
smoothing scale is $k_{L} = {\mathrm Mpc}^{-1}$ (see Fig. \ref{F1}).}
\label{F2}
\end{figure}
The first interesting exercise, for the present purposes, is reported in Fig. \ref{F2} where, in the left column, the 
power spectra of the brightness perturbations are illustrated for a wave-number $ k = 0.1\,\,{\mathrm Mpc}^{-1}$; in the right column 
the power spectra of the same quantities are illustrated for $k = 0.2\,\,{\mathrm Mpc}^{-1}$.
Concerning the results reported in Fig. \ref{F2} different comments are in order:
\begin{itemize}
\item{} for $\varepsilon = 0.1$ and $n_{r} =0.951$, the SW plateau imposes $B_{L} < 1.14 \times 10^{-8} \,\,{\mathrm G}$; from Fig. \ref{F2} it follows that a magnetic field of only $30\,\,{\rm nG}$ (i.e. marginally incompatible with the 
SW bound) has a large effect on the brightness perturbations as it can be argued by comparing, in Fig. \ref{F2}, the 
dashed curves (corresponding to $30\,\,{\rm nG}$ ) to the full curves which illustrate the case of vanishing 
magnetic fiels;
\item{} the situation where $B_{L} > {\mathrm nG}$ cannot be simply summarized by saying that the amplitudes of the power spectra 
get larger since there is a combined effect which both increases the amplitudes and shifts slightly the phases of the oscillations;
\item{} from the qualitative point of view, it is still true that the intensity oscillates as a cosinus, the polarization as a sinus;
\item{} the phases of the corss-correlations are, comparatively, the most affected by the presence of the magnetic field.
\end{itemize}
\begin{figure}
\begin{center}
\begin{tabular}{|c|c|}
      \hline
      \hbox{\epsfxsize = 7.6 cm  \epsffile{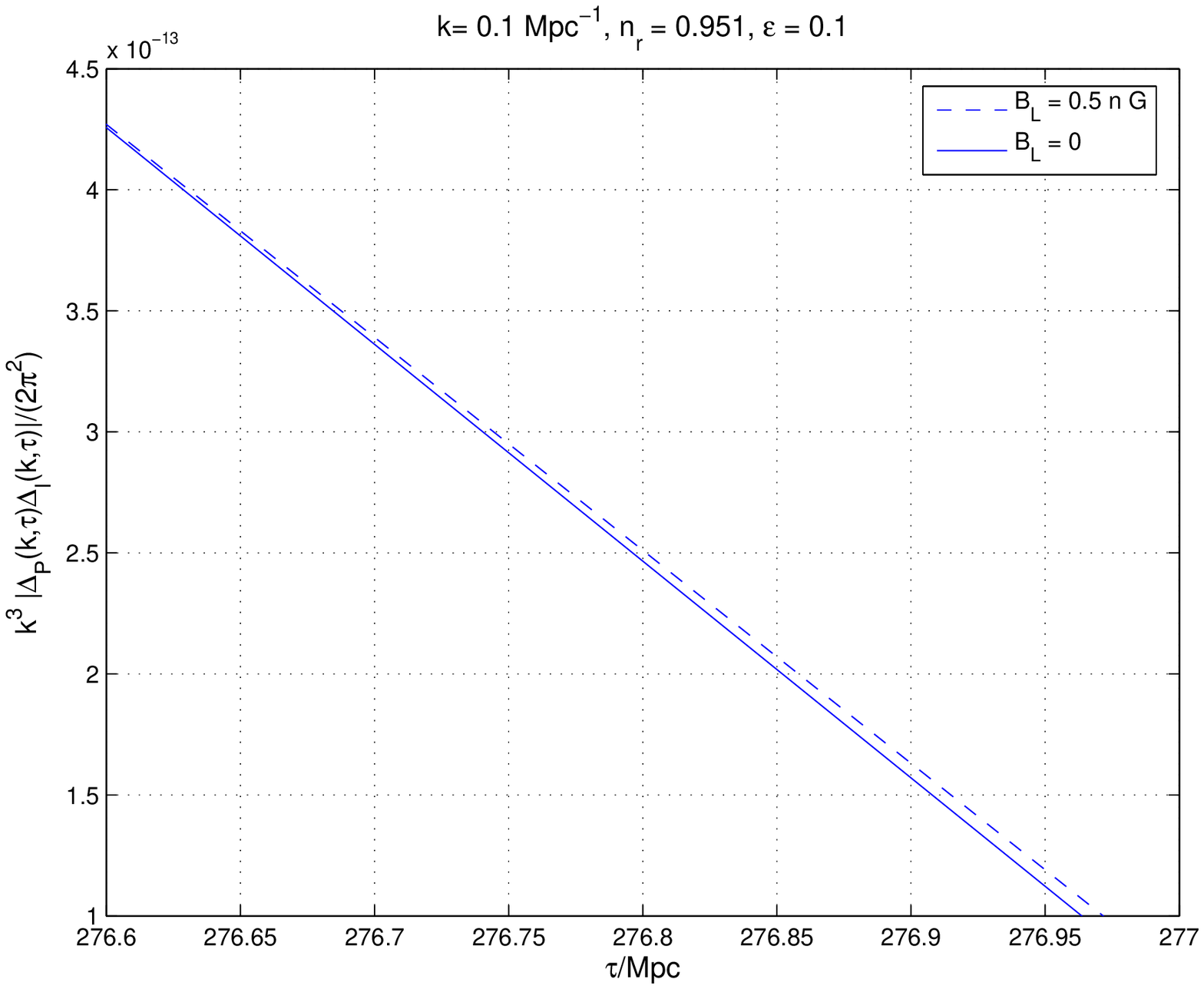}} &
      \hbox{\epsfxsize = 7.6 cm  \epsffile{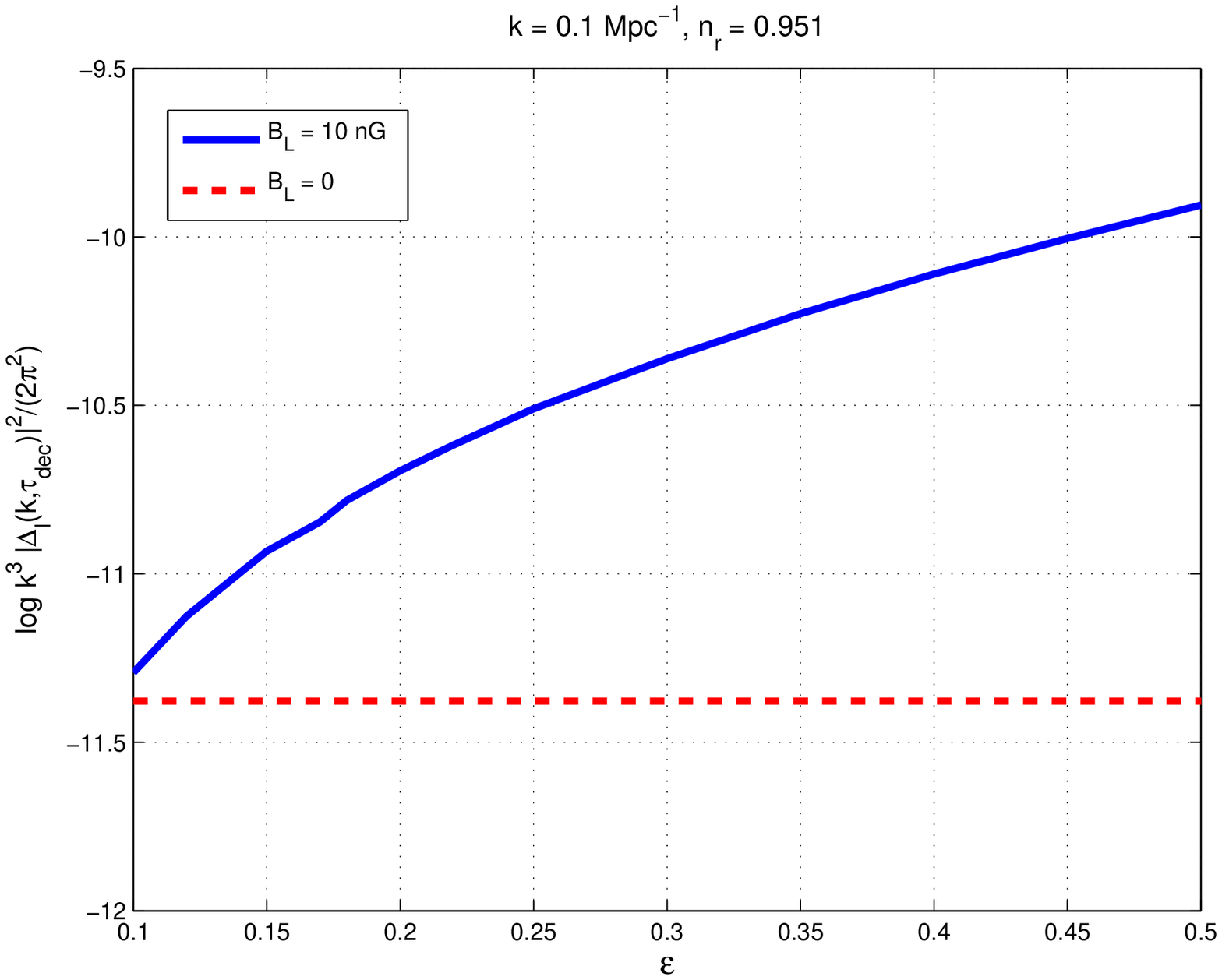}}\\
      \hline
\end{tabular}
\end{center}
\caption[a]{A detail of the cross-correlation (left plot). The autocorrelation of the intensity at $\tau_{\rm dec}$ as a function 
of $\varepsilon$, i.e. the magnetic spectral index  (right plot).}
\label{F3}
\end{figure}
The features arising in Figs. \ref{F1} and \ref{F2} can be easily illustrated 
for other values of $\epsilon$ and for different choices of the pivot or smoothing 
scales. The general lesson that can be drawn is that 
the constraint derived only by looking at the SW plateau are only a necessary condition on the strength of the magnetic 
field.  They are, however, not sufficient to exclude observable effects at smaller scales. This aspect is illustrated in the plot at the left in Fig. \ref{F3} which captures 
a detail of the cross-correlation. The case when $B_{L} =0$ can be still distinguished from the case $B_{L} = 0.5\,\,{\mathrm nG}$. 
Therefore, recalling that for the same choice of parameters the SW plateau implied that $B_{L} < 11.4\,\,{\mathrm nG}$, 
it is apparent that the intermediate scales lead to more stringent conditions even for nearly scale-invariant spectra 
of magnetic energy density. For the range of parameters of Fig. \ref{F2} we will have 
that $B_{L} < 0.5\,\,{\mathrm nG}$ which is more stringent than the condition 
deduced from the SW plateau by, roughly, one order of magnitude. 

If $\varepsilon$ increases to higher values (but always with $\varepsilon <0.5$)
by keeping fixed $B_{L}$ (i.e. the strength of the magnetic field smoothed 
over a typical length scale $L = 2\pi/k_{L}$) the amplitude of the brightness perturbations gets larger in comparison 
with the case when the magnetic field is absent. This aspect is illustrated in the right plot of Fig. \ref{F2} where the logarithm (to base $10$) of the intensity autocorrelation is evaluated at a fixed wave-number (and at $\tau_{\rm dec}$) as a function of 
$\varepsilon$. The full line (corresponding to a $B_{L} = 10\,\,{\mathrm nG}$) is progressively divergent 
from the dashed line (corresponding to $B_{L} =0$) as $\varepsilon$ increases.  

\begin{figure}
\begin{center}
\begin{tabular}{|c|c|}
      \hline
      \hbox{\epsfxsize = 7.5 cm  \epsffile{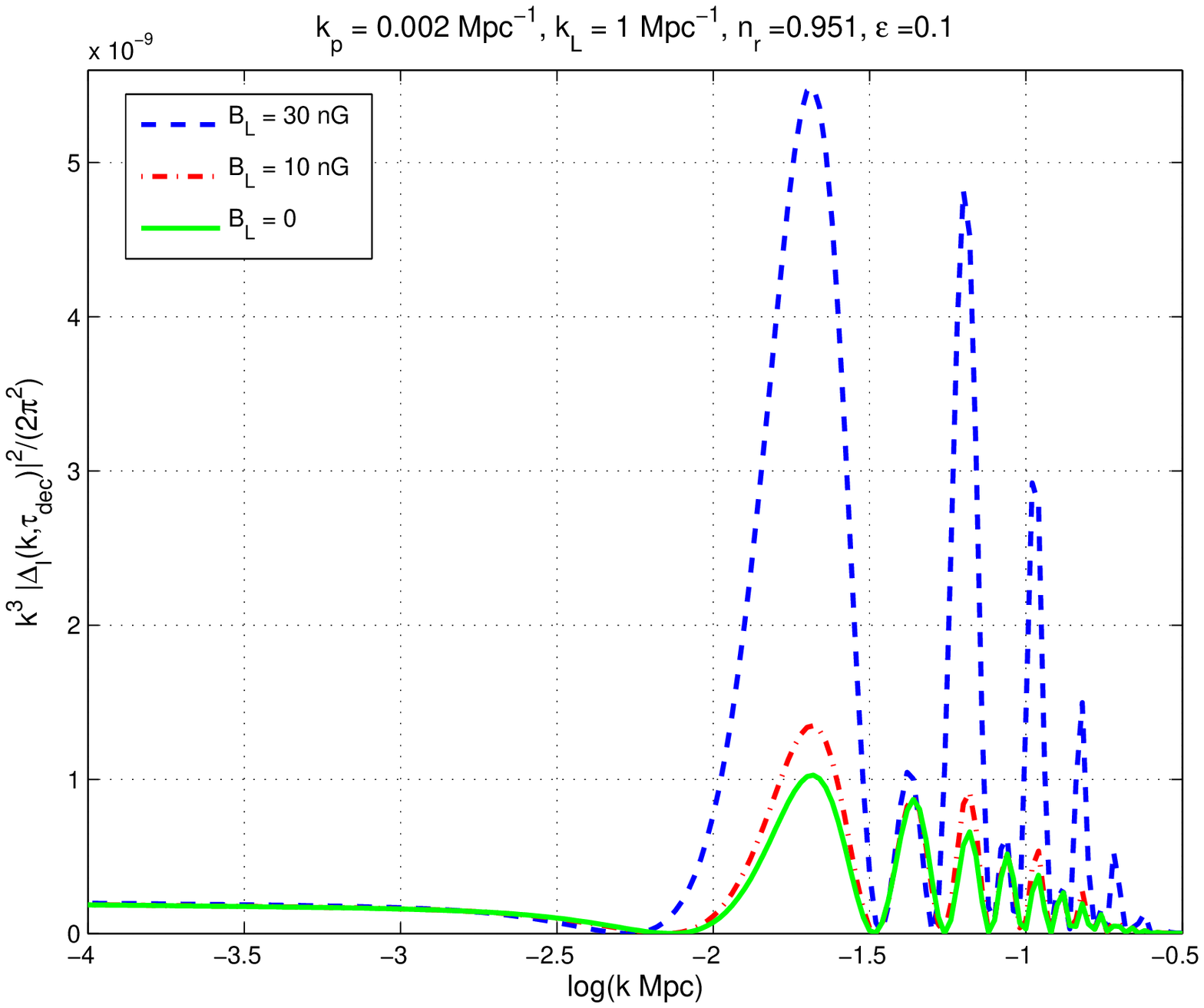}} &
      \hbox{\epsfxsize = 7.5 cm  \epsffile{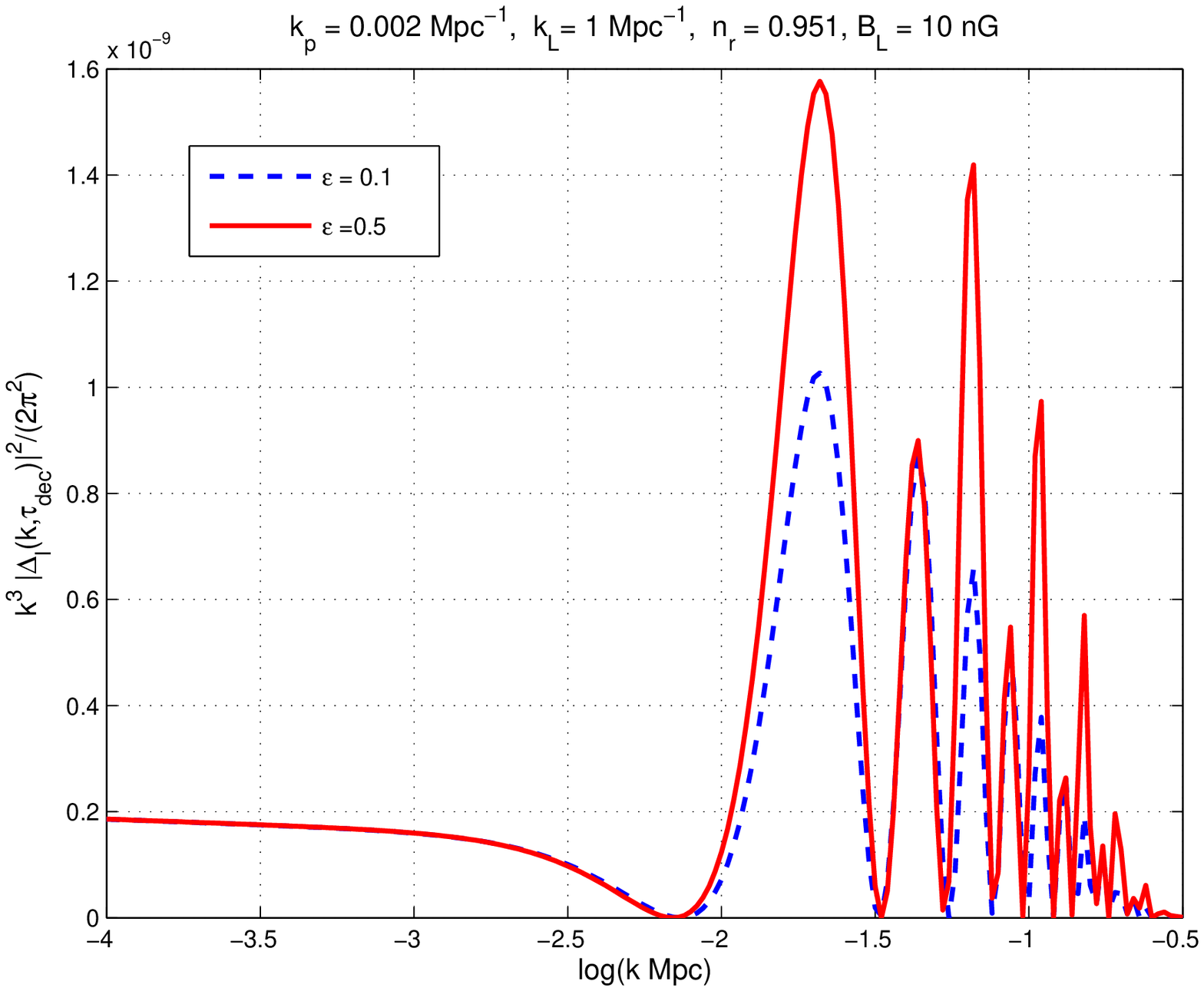}} \\
      \hline
      \hbox{\epsfxsize = 7.5 cm  \epsffile{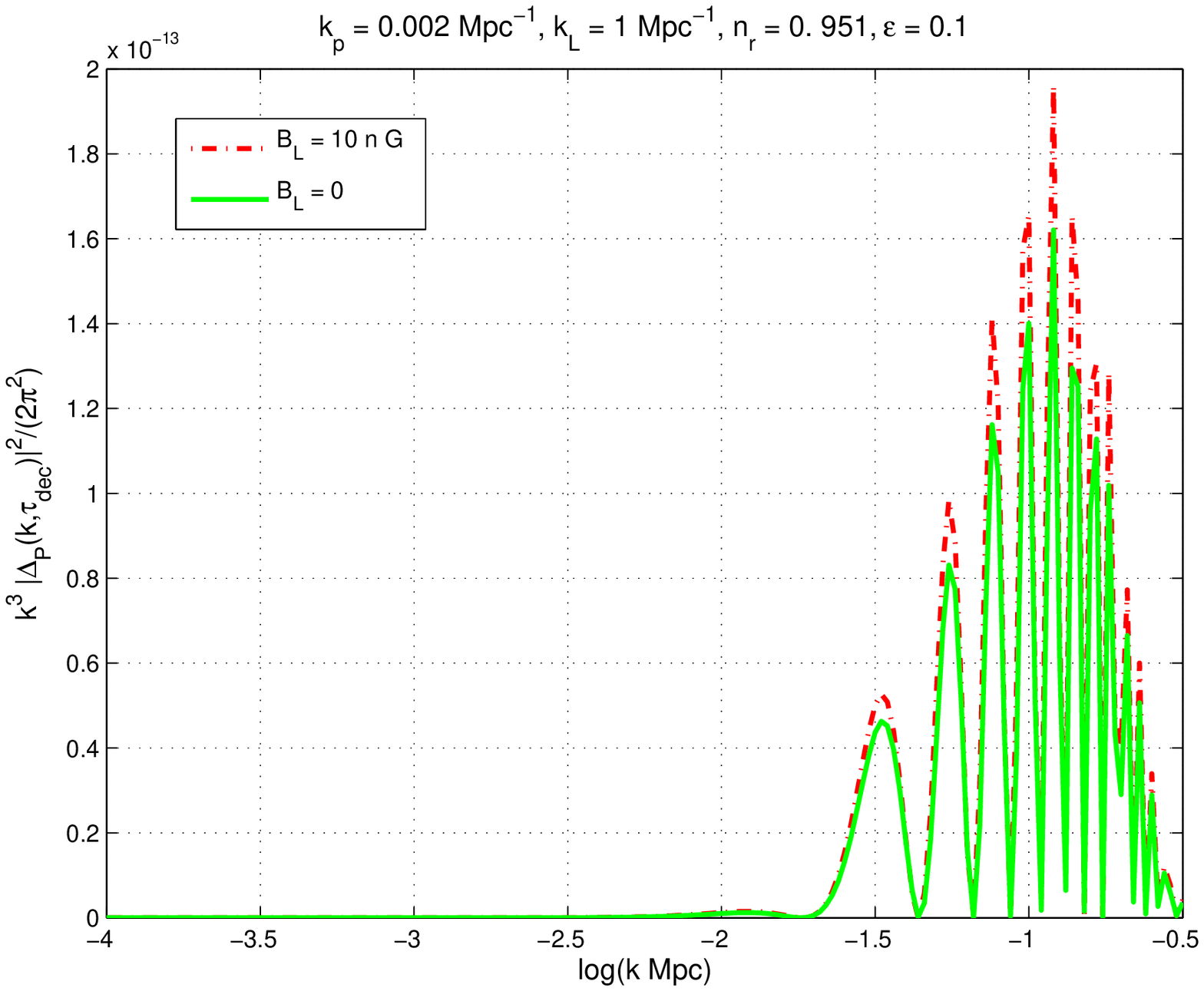}}  &
      \hbox{\epsfxsize = 7.5 cm  \epsffile{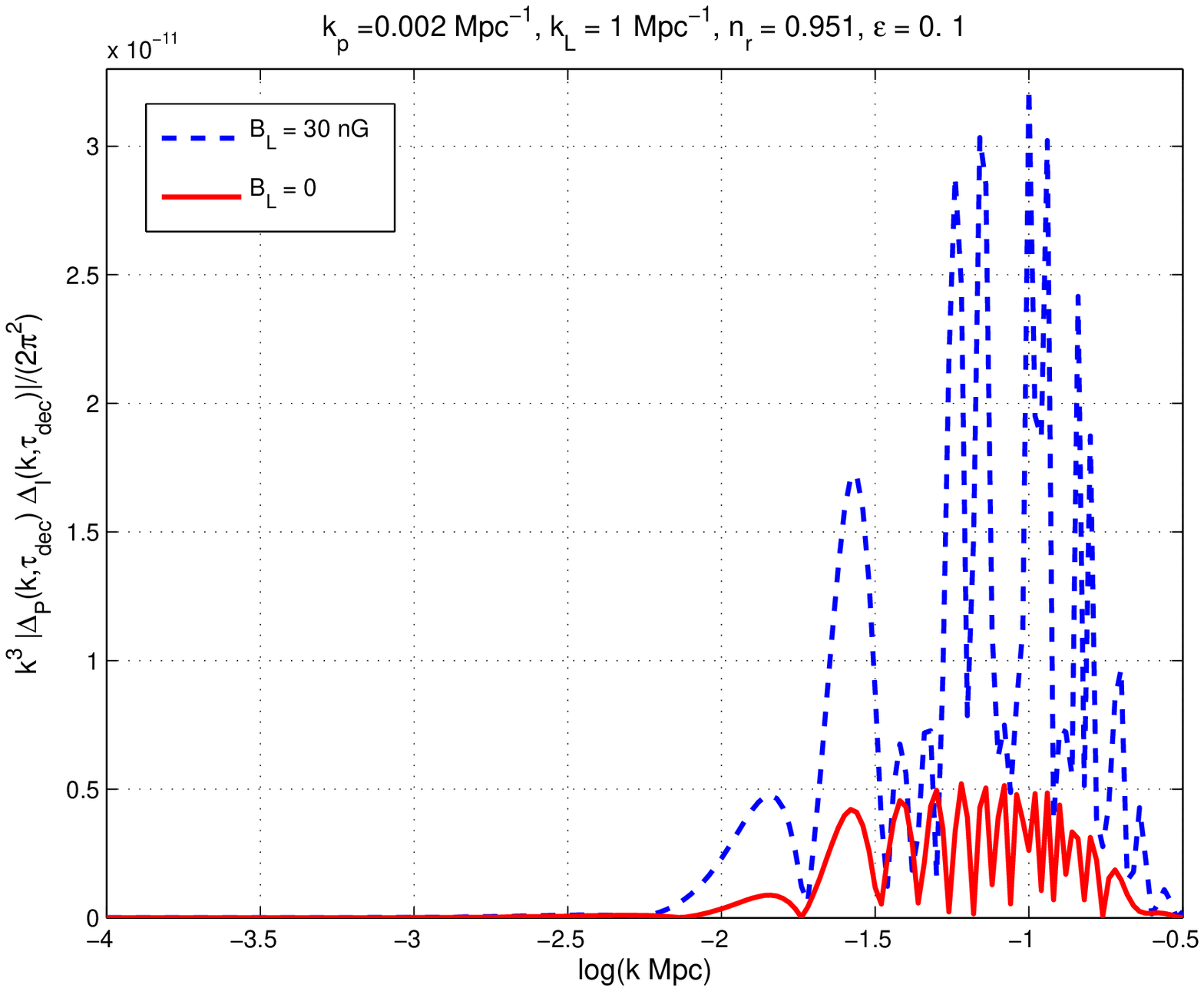}}\\
      \hline
\end{tabular}
\end{center}
\caption[a]{The power spectra of the brightness perturbations at $\tau_{\rm dec}$ for the 
parameters reported in the legends.}
\label{F4}
\end{figure}
In Fig. \ref{F4} the power spectra of the brightness perturbations are reported at $\tau_{\rm dec}$ and as a function of $k$.
In the two plots at the top the autocorrelation of the intensity is reported for different values of $B_{L}$ (left plot) and 
for different values of $\varepsilon$ at fixed $B_{L}$ (right plot). In the two plots at the bottom the polarization 
power spectra are reported always at $\tau_{\rm dec}$ and for different values of $B_{L}$ at fixed $\varepsilon$.
The position of the first peak of the autocorrelation of the intensity is, approximately,  $k_{\rm d}
 \simeq 0.017 \,\,{\mathrm Mpc}^{-1}$. 
 The position of the first peak of the cross-correlation is, approximately, 
 $3/4$ of  $k_{\rm d}$. From this consideration, again, we can obtain that $B_{L} < 0.3 \,\, {\mathrm nG}$ which is more constraining 
 than the SW condition.
 
Up to now the adiabatic mode has been considered in detail. We could easily add, however, non-adiabatic modes that 
are be partially correlated with the adiabatic mode. It is rather plausible, in this situation, that by 
adding new parameters, also the allowed value of the magnetic field may increase.
Similar results can be achieved by deviating from the assumption that the magnetic field and the 
curvature perturbations are uncorrelated. This aspect can be understood already 
from the analytical form of the SW plateau (\ref{SWP}). 
If there is no correlation 
between the magnetized contribution and the adiabatic contribution, i.e. 
$\gamma_{b r} =\pi/2$, the SW plateau will be enhanced in comparison 
with the case when magnetic fields are absent.  The same situation 
arises when the two components are anti-correlated (i.e. $\cos{\gamma_{br}}<0$).
However, if the fluctuations are positively correlated (i.e.  
$\cos{\gamma_{b r}}>0$) the cross-correlation adds negatively to the 
sum of the two autocorrelations of ${\cal R}$ and $\Omega_{\rm B}$ so that 
the total result may be an  overall reduction of the power  with respect
to the case  $\gamma_{br} =\pi/2$.

\renewcommand{\theequation}{5.\arabic{equation}}
\section{Concluding remarks}
\setcounter{equation}{0}
In the recent past, various attempts have been made to constrain large-scale magnetic fields 
from CMB physics. While interesting results have been obtained for the vector and tensor 
modes of the geometry, the scalar modes have been, comparatively, less studied.  
In this study  a systematic strategy to tackle this problem has been proposed and developed. The basic inspiration 
of the present approach can be summarized  by stating that 
the effects of fully inhomogeneous magnetic fields on CMB 
anisotropies is not univocal.  In the standard lore 
the only relevant parameters to define the effects of large-scale 
magnetic fields on CMB anisotropies are the ones related to the 
magnetic field itself. For instance, if the magnetic field is uniform 
the only parameter will be its intensity; if the magnetic field is 
fully inhomogeneous the only two parameters will be the spectral 
amplitude and the spectral slope.  
The point of view of the present investigation is that this 
way of thinking is  ambiguous and, to some extent,  not so productive.
In fact, the spectral parameters of the magnetic field are certainly 
a necessary ingredient of the analysis. They are, however, not 
sufficient to conduct the calculation of the brightness perturbations.
The information on the spectral parameters must be 
complemented by the specific pre-decoupling initial conditions. 
According to the results of this investigation, a sound  
procedure to be followed can be summarized as follows:
\begin{itemize}
\item{} include the magnetic fields in the system of scalar modes 
before matter radiation equality;
\item{} solve the system in the long wave-length approximation 
prior to decoupling;
\item{} use the obtained solutions for the numerical evaluation 
of the brightness perturbations, for instance, in the tight coupling 
approximation.
\end{itemize}
The ambiguity mentioned above arises during the first step 
of the procedure we just outlined. In fact, the inclusion 
of magnetic fields prior to equality does not lead to a single 
solution but to different solutions. We will then have 
quasi-adiabatic and quasi-entropic modes where 
a magnetized contribution can be accommodated. 

In some sense, the present proposal is more 
conservative than previous attempts: the signature of magnetic fields
on CMB physics is not directly accessible but it is mediated by the 
(supposedly rich) parameter space of the pre-decoupling initial conditions of CMB anisotropies. 
This approach is indeed common when  mixtures of adiabatic and non-adiabatic modes are studied in the absence of magnetic fields. In the latter case 
the philosophy is to specify what kind of non-adiabatic mode we want to study and constrain. Here the situation is similar and we do hope, that, in the future, the considerations reported here will serve 
as an inspiration for analyzing the effects of fully inhomogeneous magnetic fields on the scalar CMB anisotropies.

This analysis reported in this paper allowed to extend the tight-coupling approximation 
to the case when large-scale magnetic fields participate to the dynamics of the plasma.  
The evolution equations, in the tight coupling approximation, have then been integrated numerically
using, as normalization, the analytical estimate of the Sachs-Wolfe plateau modified by the presence of large-scale magnetic fields.
The reported results allow not only to set constraints 
but also to include a magnetized contribution in the current strategies of parameter extraction.  

From the combined analysis of the 
Sachs-Wolfe plateau and of the phases of Sakharov oscillations at intermediate scales, a fully inhomogeneous magnetic field uncorrelated with the adiabatic mode can be constrained to be smaller than $0.5$ nG for a slightly blue spectral slope $\varepsilon =0.1$ and for typical comoving wave-number of $1\,\, {\mathrm Mpc}^{-1}$. The conditions implied by the SW plateau ssem to be  necessary but not 
sufficient since magnetic fields compatible with the SW bound may alter Sakharov oscillations at smaller length-scales. In the most general situation, correlation angles may proliferate: there may be 
correlations between the magnetized contribution and the adiabatic or even non-adiabatic modes.  In this case 
the bounds on large-scale magnetic fields become much weaker and sizable magnetic fields may be allowed. 

The present analysis also shows that magnetic fields can affect the process of formation of CMB polarization. For instance, the usual approach to the Faraday rotation of CMB polarization is to assume that the polarization itself is only due to the adiabatic mode while the rotation of the polarization plane is only due to the magnetic field. This picture has been complemented by the
 observation that the magnetic fields affect the pre-decoupling initial conditions and, consequently,  modify the evolution of the degree of polarization. 

The tight coupling expansion is rather useful to discuss 
the integration of brightness perturbations in some simple cases.
Moreover, it is an essential ingredient of Boltzmann codes 
since it is used, at early times, to avoid the integration 
of the (stiff) Euler equations. It will be interesting, for future studies, to 
analyze the space of magnetized initial conditions by solving 
directly the whole Boltzmann hierarchy.

\end{document}